\documentclass[final,5p,times,twocolumn]{elsarticle}

\usepackage{amssymb}

\usepackage{makecell}
\usepackage{microtype}
\usepackage{caption}
\usepackage{subcaption}

\usepackage[export]{adjustbox}
\usepackage{graphicx}
\usepackage{amsfonts}
\usepackage{gensymb}
\usepackage{amsthm}
\usepackage{amsmath}
\usepackage{booktabs}
\usepackage{multirow}

\usepackage{algorithm}
\usepackage{algpseudocode}

\usepackage[labelfont=bf,justification=raggedright,singlelinecheck=true,tableposition=top]{caption}
\usepackage{prettyref}
\usepackage{varioref}
\usepackage{cleveref}

\usepackage[toc]{appendix}

\newtheorem{assumption}{Assumption}

\newrefformat{tab}{Table~\ref{#1}}
\newrefformat{fig}{Fig.~\ref{#1}}
\newrefformat{section}{Section~\ref{#1}}
\newrefformat{eqn}{Eq.~(\ref{#1})}

\newcommand{\gt}{>}
\newcommand{\lt}{<}
\newcommand{\R}{\mathbb{R}}
\newcommand{\N}{\mathbb{N}}
\newcommand{\argmax}{\arg\max}
\newcommand{\argmin}{\arg\min}

\algblock{Initialize}{EndInitialize}
\algnotext{EndInitialize}

\begin{document}

\begin{frontmatter}



\title{Formation control with connectivity assurance for missile swarm: a natural co-evolutionary strategy approach}

 \author[address1]{Chen Junda}
 \ead{CJD@e.gzhu.edu.cn}

\address[address1]{School of Mechanical and Electrical Engineering, Guangzhou University, Guangzhou  {\rm 510006}, China}

%
%

\begin{abstract}

Formation control problem is one of the most concerned topics within the realm of swarm intelligence, which is usually solved by conventional mathematical approaches. In this paper, however,  we presents a metaheuristic approach that leverages a natural co-evolutionary strategy to solve the formation control problem for a swarm of missiles. 

The missile swarm is modeled by a second-order system with heterogeneous reference target, and exponential error function is made to be the objective function such that the swarm converge to optimal equilibrium states satisfying  certain formation requirements. Focusing on the issue of local optimum and unstable evolution, we incorporate a novel model-based policy constraint and a population adaptation strategies that greatly alleviates the performance degradation.  With application of the Molloy-Reed criterion in the field of network communication, we developed an adaptive topology method that assure the connectivity under node failure and its  effectiveness are validated both theoretically and experimentally.
 Experimental results valid the effectiveness of the proposed formation control approach. More significantly, we showed that it is feasible to treat generic formation control problem as Markov Decision Process(MDP) and solve it through iterative learning.
\end{abstract}


\end{frontmatter}


\section{Introduction}
\label{section:introduction}
Intelligent control of swarm system has being widely used in tracking, rescue and even delivery \cite{limFormationControlLeader2009, shiSurveyIntelligentControl2021}. Through the interactive cooperation of multiple agents, the swarm can exhibit complex intelligent behaviors, such as cohesion, separation and alignment, which are know as Reynolds rules \cite{reynoldsFlocksHerdsSchools}. Among the various research directions, the formation control problem has been widely studies on a variety of system models for its unique application value and complexity.

In this paper, we focus on the formation control of missile swarm. Similar to common formation control problems, the formation control for missile swarm covers formation initialization, contraction, expansion and reconfiguration \cite{naigangcuiResearchMissileFormation2009}, corresponding with formation producing and formation tracking problems according to \cite{renDistributedCoordinationMultiagent2011}. Based on the sensing capability and interaction level, the current dominant research in  \cite{ohSurveyMultiagentFormation2015} classifies formation control methods into position-, displacement-, and distance-based control. The main difference between the different control methods is the ability to sense relative or absolute state information.  For this reason, we propose to classify angle-based control as a special case of displacement-based control, in which the relative distance constraint from the relative position is removed. In this paper, we focus on the displacement-based control framework, not only for its simplicity and stability, but also for its huge realistic application value outside of cyclic pursuit problem \cite{marshallFormationsVehiclesCyclic2004}. 

In practice, when dealing with large-scale swarm systems, control methods that the system can adopt are usually subject to limited communication bandwidth, communication quality and other interference issues, so more flexible communication means that are tolerant of communication failures are necessary for achieving robust information transmission. However, traditional communication methods based on fixed communication topology are difficult to cope with unexpected situations and provide continuous and reliable communication. For developing more robust intelligent swarm communication algorithm, two newly developed adaptive communication methods in the field of networked control system are known as \emph{ad-hoc} based network and cluster based network \cite{asaamoningDroneSwarmsNetworked2021}. 
The author in \cite{chenRobustIntelligentDrone2020} developed a more robust intelligent swarm communication algorithm based on the Molloy-Reed criterion and grey-prediction, improving the robustness of the drone swarm network and the reliability of data transmission. Such algorithm is further employed in flying \emph{ad-hoc} network (FANET), which is a distributed and self-organizing communication framework.  Under the same \emph{ad-hoc} framework, a new approach is developed utilizing received signal strength instead of localization facilitates to measure individual distances \cite{shritNewApproachRealize2017}. Considering that distance is usually closely related to communication quality, and that the perception of distance does not require an additional communication bandwidth, it is highly feasible to regard distance as a major factor in configuring the communication topology. In this work, we develop an adaptive topology communication method based on minimizing the communication distance and overcome the problem of head failure in cluster-based communication method to ensure the connection stability for formation control.
Formation control methods based on traditional control theory and dynamic systems can provide more robust control for single control objectives and motion patterns \cite{slotineAppliedNonlinearControl1991}, but it is hardly to achieve high precision coordination through conventional control methods with multiple objectives or in environments that are too complex to be modeled. That is why intelligent control approaches are introduced as more flexible alternatives for operating in unstructured or dynamic environments surrounded with multiple uncertainties. 

A number of metaheuristic algorithms have been shown to cope well with multi-objective complex constrained optimization problems and have been widely used in all aspects of formation control, including but not limited to formation optimal configuration, motion planning, etc. 
A comprehensive survey on the development of such algorithms for aircraft motion planning problem is presented in \cite{wuSurveyPopulationbasedMetaheuristic2021}. In \cite{liuOptimalDesignMultimissile2021} Liu investigated a PSO-based algorithm in the optimal design of missile formation configuration. In \cite{seung-mokleeCooperativeCoevolutionaryAlgorithmBased2015}, Seung-Mok propose a cooperative co-evolution PSO algorithm based on the traditional PSO-based MPC, which optimize in a distributed way and improves the speed and performance of the original algorithm. Another population-based metaheuristic known as the genetic algorithm (GA) is employed to evolve positioning strategy of a formation controlled multi-robot systems \cite{pessinIntelligentControlEvolutionary2010}. A gradient-descent-based reinforcement learning method utilizing actor-critic framework is proposed for optimal consensus control of multi-agent systems. Existing research on consensus control provides another perspective on solving the formation problem, since general formation control problem is a special kind of consensus problem which requires that certain errors are maintained between the states of neighboring robots, rather than identical states.
Although existing metaheuristic-based algorithms have been applied to formation control to search for single-step optimal solutions in real time, these methods are usually slow when running and do not have the ability to migrate and learn. In contrast, neural network-based controllers can be trained to achieve the same performance through iterative learning, and can be easily deployed to compact mobile units with weak computational performance and low cost despite of the majority of the computational cost spent at the training stage.

There is a dearth of existing research using neural network controllers, despite the fact that they are commonly used under the reinforcement learning (RL) paradigms \cite{nguyenDeepReinforcementLearning2020} and trained to solve specific control or decision making problems.  In \cite{liAdaptiveSOMNeural2018}, Liu developed a SOM-based neural network approach for motion planning in the formation control problem of a multi-AUV system. By the way, the SOM is a sort of unsupervised neural network that usually serves for specialized purpose, like, competitive learning and task assignment \cite{barretoIdentificationControlDynamical2004}. The an iterative algorithm to find the most energy-sufficient motion position for each AUV. The algorithm is an iterative algorithm that is used to search for the most energy-efficient motion path of the agent. Since the algorithm takes all individuals’ states as input and the target path as output, it can be regarded as centralized method which is not desirable for the practical deployment requirement.

The general idea of the heuristic algorithm using neural network controller is similar to other algorithms such as PSO, ACO, etc., which are designed to find the optimal solution of the cost function under the constraints by iterative search, but the most essential difference between the former and the other algorithms is that the former optimizes the neural network weights and the final optimized controller can be deployed directly, while the latter optimizes directly in the solution space by single-step. Both methods have their advantages and disadvantages. Since the latter does not need pre-training, it relies too much on single-step computation and it is difficult to guarantee the speed requirement for real-time operation. Therefore neural network controller as an adaptive and learnable controller is considered as a promising means of intelligent control in the future  \cite{zhangDataDrivenOptimalConsensus2017}.

Apart from acting as a controller to generate control commands, neural networks are widely used in control systems for sensing, decision making, trajectory planning, and many other purposes. In \cite{wangCooperativeUAVFormation2007}, a modified Grossberg neural network (GNN) is used to generate the shortest path to avoid the obstacle and reach the target point. Lan adopts an RBFNN to estimate the system disturbance in \cite{lanAdaptiveNeuralNetworkBasedShapeControl2018}, which leads to enhanced robustness and adaptability of swarm formation control based on the artificial potential field method. Additional researches that implement adaptive neural networks to estimate the uncertain and nonlinear dynamics of the system can be viewed in \cite{feiNeuralNetworkAdaptive2020, niAdaptiveNeuralNetwork2021, yangNeuralnetworkbasedFormationControl2021}. Further, in \cite{lanCooperativeControlSwarming2020}, Lan use reinforcement learning theory to train a neural network controller that can be applied to the distributed control of swarm system in an unknown dynamic environment, where the agents in the swarm can perform basic intelligent behaviors such as tracking and obstacle avoidance. Likewise, extensive research has shown that neural networks have certain robustness in many scenarios without inferior to traditional methods, and are relatively more flexible and applicable.

To investigate the applicability of neural network controllers within the framework of formation control problem, in this paper we propose to adopt a multi-layer perceptron (MLP) neural network controller using a cooperative NES-based approach and confirm that the formation control problem can also be solved as a multi-agent collaborative task by the neural network controller. 

The foremost motivation of this paper is to develop a metaheuristic evolutionary computational approach to solve the formation control problem for MASs, in the meanwhile exploring the usage of neural networks(NNs) controller. We use a recently proposed NCES algorithm \cite{chenCooperativeGuidanceMultiple2022} to realize such of a vision and apply it to the control of a second-order multi-missile system. We incorporate a policy constraint approach to enhance the stability of the algorithm so as to optimize the objective function to find the Nash equilibrium strategy. The proposed algorithm is more flexible compared with conventional approaches \cite{naigangcuiResearchMissileFormation2009, xingguangTimevaryingFaulttolerantFormation2019, weiOptimalFormationKeeping2012, zhangNovelCooperativeControl2021}. In addition, an adaptive topology scheme is designed to solve the common node failure problem in formation control, and this method can achieve stable communication connections at relative low communication cost. To further improve the performance of the algorithm and mitigate the local optimum issue, we also propose a stable population adaptation method. Emulation experiments show that the proposed formation control algorithm can handle tasks such as formation maintenance , reference trajectory tracking and formation switching in the face of obstacles with high accuracy, and is immune to disturbances such as random initial position and node failure.

The remainder of this work continues as follows.  In \prettyref{section:Preliminaries and Problem Formulation}, the non-linear multi-missile system is modeled and the displacement-based formation control problem is formulated, including the specification of formation patterns, while the background of NCES algorithm is also introduced in this section. In \prettyref{section:Implementation of Natural Co-evolutionary Strategy based formation control via neural networks}, an distributed NCES-based formation control algorithm using a neural network controller is proposed, and in \prettyref{section:Simulation and result analysis} experiments are conducted for a variety of scenarios and results are derived. Finally, conclusions are presented in \prettyref{section:Conclusion}.

\section{Preliminaries and Problem Formulation}
\label{section:Preliminaries and Problem Formulation}
\subsection{System modeling of swarm of cruise missiles}
\begin{figure}[htbp]
	\centering
	\includegraphics[scale=0.3]{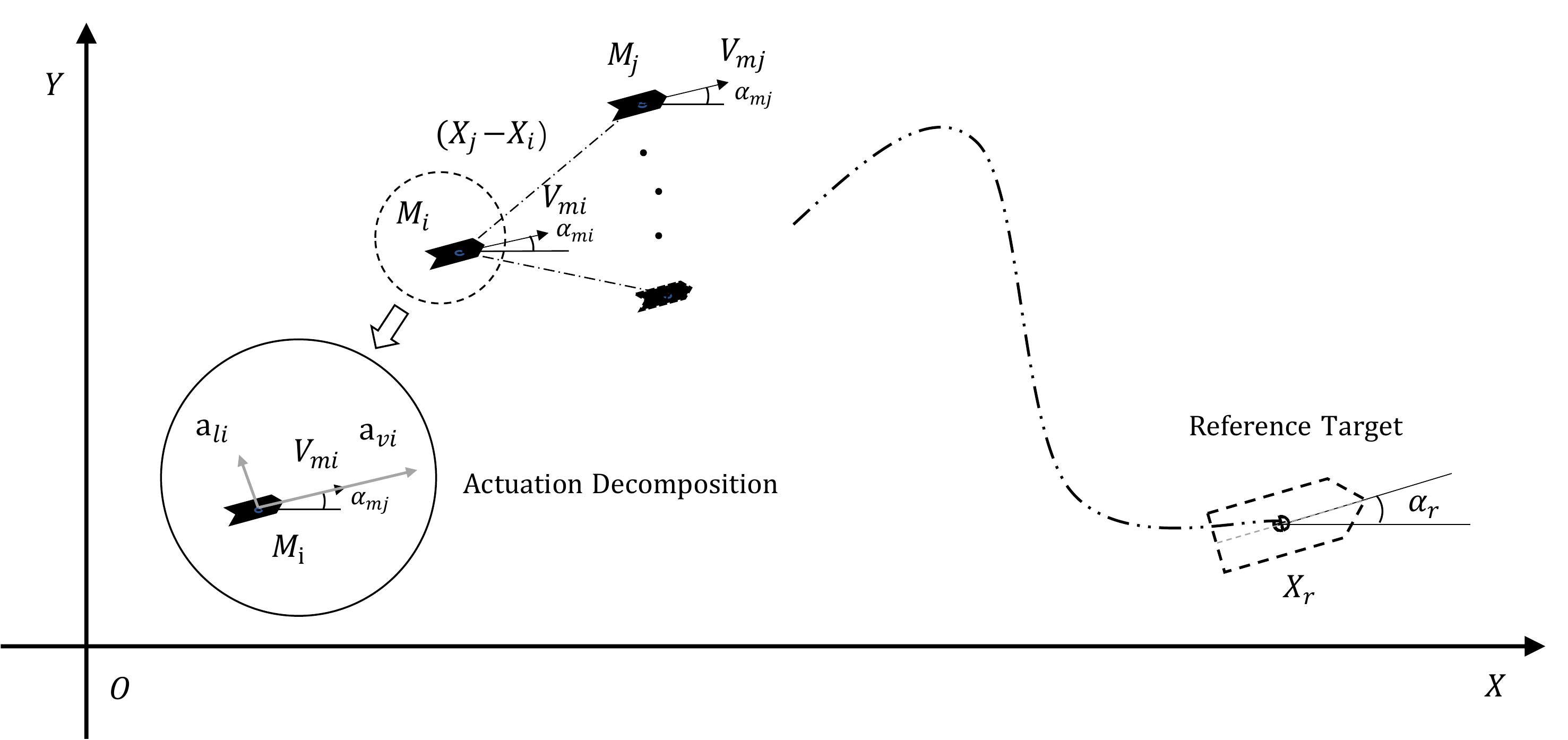}
	\caption{engagement scenario}
	\label{fig:engagement1}
\end{figure}

This paper focus on the formation control in two-dimensional space, i.e., the $OXY$ planar space of the inertial coordinate frame. As show in \prettyref{fig:engagement1}, the swarm is consist of multiple missiles, each missile can be treated as point mass, In the simplified dynamics model we do not consider aerodynamic factors and the effect of the missile's own inertia tensor. Consider a total of $N$ missiles, for missile $i$  denoted by $M_i$, $V_{mi}, \alpha_{mi}$ $x_{mi}$ and $y_{mi}$ are used to represent the velocity and heading angle, and its coordinates in the global coordinate frame, respectively. Let $X_i = [x_{mi}, y_{mi}, \alpha_{mi}]^T$ represents the state vector of the missile, which are not necessarily required hereinafter, and $X_r = [x_{r}, y_{r}, \alpha_{r}]^T$ represents the state of the reference target. For brevity, let $\hat{X}_i=[X_i^T,V_{mi}]^T$ be the extended state vector with speed inserted. As shown in the partial enlargement of the actuation decomposition, the missiles have two independent input $a_{vi}, a_{li}$ which are the acceleration commands both along and perpendicular to the direction of the velocity, respectively.
The second-order system dynamic of the $i^{th}$ missile can be expressed as
\begin{equation}\label{eqn:systemDynamic}
	\begin{cases}
		\dot{\hat{X}}_i=\Psi(V_{mi},\alpha_{mi}) + \Phi(V_{mi})u_i\\
		X_i = A\hat{X_i}\\
	\end{cases},
\end{equation}

where $u_i=[a_{vi}, a_{li}]^T$ is the input for $M_i$, and

\begin{equation}
	\begin{aligned}
		\Psi(V_{mi}, \alpha_{mi})
		&= \begin{bmatrix}
			V_{mi}\cos\alpha_{mi}\\
			V_{mi}\sin\alpha_{mi}\\
			0\\
			0
		\end{bmatrix}, \\
	\Phi(V_{mi})&=\begin{bmatrix}
			0 & 0\\
			0 & 0\\
			0 & 1/V_{mi}\\
			1 & 0
		\end{bmatrix},\\
	 A&=\begin{bmatrix}
			I_3, 0_{3,1}
		\end{bmatrix},
	\end{aligned}
\end{equation}
where $I_n, 0_{m,n}$ denote the identity and zero matrix, and $j$ represents the neighboring missile. Note that system \prettyref{eqn:systemDynamic} is similar to the system discussed in \cite{aicardiClosedLoopSteering1995} which is often called unicycle-like vehicle. In this paper, the missile is a nonholonomic system with controllable speed and capable of maneuvering with limited actuation and dynamic constrains. We assume that under certain communication relationship, the relative state $z_{ji}$ can be sensed by missile $i$. In addition we make the following assumptions about the system:
\begin{assumption}
	We assume that the reference target/leader is a first-order model driven by unknown changeable speed and angular velocity commands, which means that its vertical acceleration is not available but its lateral acceleration also subject to the constraints same as the follower missiles. Hence the heterogeneous system may bring more complexity for mathematical analysis and stability issue.
\end{assumption}
\begin{assumption}
	It is assumed that missile controller applies the Cascade control \cite{dineshReviewCascadedLinear2021} structure, such that its altitude is automatically controlled by the inner-loop control, which is a stable closed-loop. We focus on the design for the missiles’ trust and lateral acceleration of the outer control loop. This divide and conquer approach is widely used in many of the literature \cite{weiOptimalFormationKeeping2012, renConsensusStrategiesCooperative2007} to simplify the problem in order to verify the effectiveness of the control method, and is also valid for three dimensional dynamic models.
\end{assumption}

\subsection{Formation control under displacement-based framework}
The formation pattern is defined by $\lambda=\{\lambda_{pi}=[\lambda_{xi}, \lambda_{yi}]^T: i=1,2,...,N\}$,  which is the set of Cartesian coordinates of the origin of each node that are expressed in the coordinate system,  such that the center of the formation is the origin which should satisfies
\begin{equation}\label{eqn:formationDefPrerequisite}
	\sum_{i=1}^N \lambda_{xi}=0, \sum_{i=1}^N \lambda_{yi}=0. 
\end{equation}

Let $\mathcal{N}_i $ denotes the set of neighboring missiles of the $i^{th}$ missile.  In the displacement-based framework, each missile has to align its own coordinate system with the global coordinate system and to be able to sense the relative positions and orientations of its neighbors, whereas its position in the global coordinate system is not necessarily required. Suppose that the reference target should be kept in the formation center, then the tracking error of the $i^{th} $ missile is 
\begin{equation}
	e_i = \sum_{j\in \mathcal{N}_i}[(P_i-P_j)-(X_i - X_j)] +  \zeta_i(X_r + P_i - X_i),
\end{equation}
where $P_i =[\lambda_{xi}, \lambda_{yi}, 0]^T$ and $\zeta_i = 1$ if the state of the target is available to the $i^{th}$ missile, else $\zeta_i = 0$, $e_i \in \mathbb{R}^3$.  It is worth noting that the first term represents the formation maintenance error, while the second term represents the target tracking error when it is able to obtain the target information.  
However, the error vector is defined in the global coordinate system, in order to standardize the effect of individual missile’s heading angles on the error vector to the relative coordinate system that are aligned with the missile’s orientation, we define the rotated error vector as $e_{ri}= R_3(-\alpha_i)e_i,$$ $ where $R_3(-\alpha)\in\mathbb{R}^{3\times3}$ is the three dimensional rotation matrix defined as
\begin{equation}\label{eqn:rotationmatrixClockwise}
	R_3(-\alpha_i) = 
	\begin{bmatrix}
		\cos\alpha_i & \sin\alpha_i & 0\\
		-\sin\alpha_i & \cos\alpha_i & 0\\
		0 & 0 & 1
	\end{bmatrix}
\end{equation}

Let’s $e_{ri}= \begin{bmatrix} e_{ix} & e_{iy} &e_{i\theta}\end{bmatrix}^T$. Considering system \prettyref{eqn:systemDynamic},  by differentiating $e_{ri}$ with respect to time and considering the discretization of system control loop,  we obtain the following error dynamic
\begin{equation}\label{eqn:errorDynamic}
	\dot{e_{ri}} = G_iu_i + \sum_{j\in \mathcal{N}_i}F_{ij}u_j + D_i u_r + H_i,
\end{equation}
where
\begin{equation}
	\begin{aligned}
		G_i &= \begin{bmatrix}-\tau(L_i+\xi_i) & e_{iy}/v_{i}\\
			0 & -e_{ix}/v_{i}\\
			0 & -(L_i+\xi_i)/v_{i}
		\end{bmatrix}, \\
		F_{ij} &= \begin{bmatrix}\tau \cos\theta_{ji} & 0\\
			\tau\sin\theta_{ji} & 0\\
			0 & 1/v_{j}
		\end{bmatrix}, \\
		D_i &= \begin{bmatrix}\xi_i\cos\hat{\theta_i} & 0\\
			\xi_i\sin\hat{\theta_i} & 0\\
			0 & \xi_i
		\end{bmatrix}, \\
		H_i &= \begin{bmatrix}-(L_i+\xi_i)v_i + \sum_{j\in \mathcal{N}_i}\cos\theta_{ji}v_j\\
			\sum_{j\in \mathcal{N}_i}\sin\theta_{ji}v_j\\
			0
		\end{bmatrix}.
	\end{aligned}
\end{equation}

$L_i$ is the number of missiles belonging to the set of neighbors $\mathcal {N}_i$ and $\tau$ is the simulation step size. Furthermore, let $u_i=[a_{vi}, a_{l}]^T$and $u_r=[v_r, w_r]^T$ denote the vector of control command, and $\tau$ be the control loop constant which is selected by the controller. 

Generally, the objective metric at time $t $ can be defined by function of the error vector, which is
\begin{equation}\label{eqn:objectiveFunction}
	J_i(t, e_{ri}(t)) =  \exp(- e_{ri}^T(t)K_Ce_{ri}(t)) ,
\end{equation}
where $K_C= \rm{diag}[k_1, k_2, k_3]$ is the symmetric positive defined matrix with positive weights.  By choosing $K_C$ properly, the swarm balances between keeping the formation shape or tracking the reference target as a priority. The optimal design of the formation controller $u_i$ can be formulated as a nonlinear optimization problem 
\begin{equation}
	u_i = \argmax_{u_i}\int_0^T J_i(t, e_{ri}(t))\text{d}t,
\end{equation}
subjected to 
\begin{equation}\label{eqn:systemConstraints}
	\begin{aligned}
		u_{min}\le u_i \le u_{max},\\
		V_{min}\le V_{mi} \le V_{max},
	\end{aligned}
\end{equation}
in which $T $ denotes the total flight time, and $u_{min}, u_{max}, V_{min}, V_{max}$ represents the system constraints.  

\begin{figure}[htbp]
	\centering
	\begin{subfigure}[h]{0.5\columnwidth}
		\centering
		\includegraphics[scale=0.8]{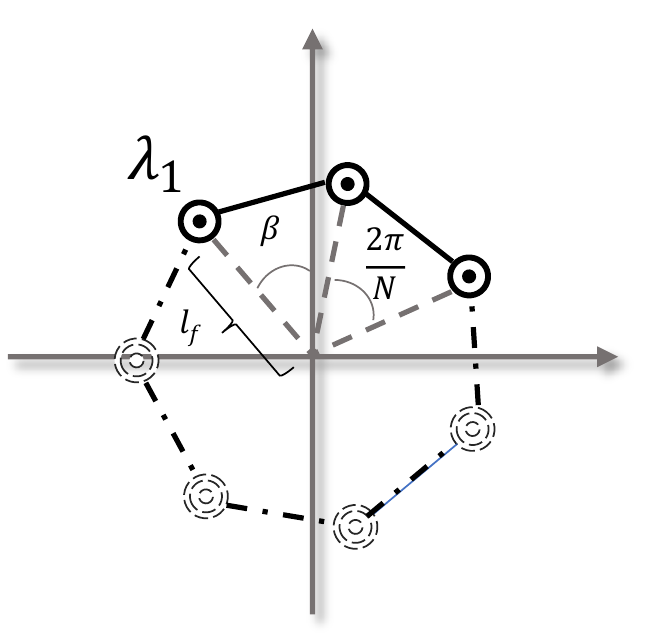}
		\caption{Regular polygon formation}
		\label{fig:regularpolygon}
	\end{subfigure}
	\begin{subfigure}[h]{0.5\columnwidth}
		\centering
		\includegraphics[scale=0.8]{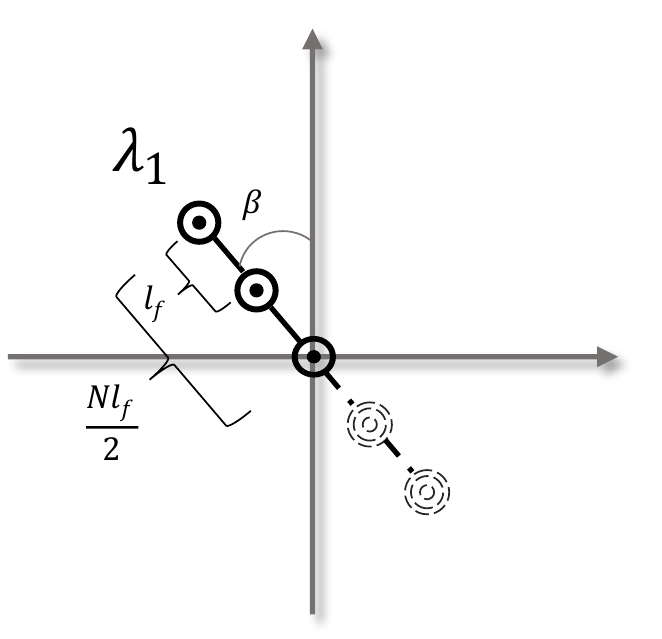}
		\caption{Straight-line formation}
		\label{fig:straightline}
	\end{subfigure}
	\caption{Two types of formation patterns}\label{fig:formationPatternsDefinition}
\end{figure}

In this paper we discuss the parametric definition of the formation geometry, which is rarely mentioned in the other works. We define two formation patterns, the regular polygon and the straight-line formation which are shown in \prettyref{fig:formationPatternsDefinition}, and denote them by $\lambda^P_{(\alpha_p,l_f)}$ and $\lambda^L_{(\alpha_p,l_f)}$.  $l_f$ and $\alpha_p$ are the parameters that control the size and rotation of the formation, specifically we have 
\begin{equation}\label{eqn:formationDef}
	\begin{aligned}
		\lambda^P_{(\alpha_p,l_f)} &= \{R_2(\alpha_p)[l_f\cos(\frac{2\pi}{N}(i-1)), l_f\sin(\frac{2\pi}{N}(i-1))]^T, \\&i\in\{1,2,...,N\} \},\\
		\lambda^L_{(\alpha_p,l_f)} &= \{R_2(\alpha_p)[l_f\frac{N-2i+1}{2}, 0]^T: i\in\{1,2,...,N \}\},
	\end{aligned}
\end{equation}
and $R_2(\beta)$ is a two-dimensional rotation matrix similar to \prettyref{eqn:rotationmatrixClockwise}, which is 
\begin{equation}
	R_2(\alpha_p) = 
	\begin{bmatrix}
		\cos\alpha_p & -\sin\alpha_p \\
		\sin\alpha_p & \cos\alpha_p \\
	\end{bmatrix}.
\end{equation}

From the above definition, it can be intuitively observed that the straight-line formation is symmetric with respect to the origin of the coordinate system that defines it, so that the sum of its $x$ and $y$ coordinates is 0, which satisfies \prettyref{eqn:formationDefPrerequisite}. Similarly when the number of nodes of a regular polygon is even, we can also get this conclusion. Still when the number of nodes is odd, we can prove that it still satisfies this prerequisite, and the following is a simple proof.

Suppose that the regular polygon formation defined in \prettyref{eqn:formationDef} consists of $N\in\{2n+1:n\in \mathbb Z\}$ nodes, the Cartesian coordinates of each node in the formation are denoted by $P_i = [\lambda_{xi}, \lambda_{yi}]^T$, so that the sum of all coordinates is
\begin{equation}
	\sum_{i=1}^{N}P_i = \begin{bmatrix}
		\xi_x \\
		\xi_y
	\end{bmatrix},
\end{equation}

Suppose the angle between the vector from the coordinate origin to an arbitrary node and the positive direction of the $Y$-axis is $\rho$. We can obtain the following sum of coordinates by multiplying with the rotation matrix to rotate node one to the $Y$-axis
\begin{equation}
	R_2(\rho)\sum_{i=1}^{N}P_i = \begin{bmatrix}
		\xi_x\cos\rho -\xi_y\sin\rho\\
		\xi_x\sin\rho +\xi_y\cos\rho
	\end{bmatrix}. \\
\end{equation}

At this configuration, since the formation is symmetric about the $Y$-axis, the sum of its x-coordinates can be obtained as $0$, so we have
\begin{equation}
	\xi_x\cos\rho -\xi_y\sin\rho = 0,
\end{equation}
thus $\xi_x = \xi_y=0$, \prettyref{eqn:formationDefPrerequisite} holds. It is shown that many other formation patterns can be considered as variants of the above two formation patterns, such as \emph{Row, Column, Square} patterns \cite{limFormationControlLeader2009}. Moreover, as discussed in the following \prettyref{section:Implementation of Natural Co-evolutionary Strategy based formation control via neural networks}, this formation paradigm can also generate some asymmetric patterns such as \emph{Wedge, Crescent} patterns \cite{lanAdaptiveNeuralNetworkBasedShapeControl2018}, etc., by deleting nodes appropriately.

\subsection{Natural Co-evolutionary strategy for MAS}
With limited sensing capability, the MAS system can be modeled as multi-agent partial observable Markov decision process(POMDP), which is an extension on the basis of the fundamental Markov decision process(MDP). Such process can be expressed as 
\begin{equation}
	P(S \times A)\to S',
\end{equation}
with $S $ and $S'$ representing the system state before and after transition, and $A=[u_1,...,u_n]$ is the actions of all agents in the system. $P(\cdot) $ is the transition probability function. It is said to be deterministic optimization problem if $P(\cdot)=1$ for all time,  otherwise it is stochastic optimization problem.

Considering of the simplified system model neglecting uncertainties, the formation control problem discussed in general can be solved using optimization algorithms under multi-agent POMDP. Assuming that the fitness function measuring the performance of agent $i$ is $f_i(\theta_i, \theta_{\mathcal{N}_i})$, in which $\theta_i $ is the controller(policy) parameter representing agent $i$, and $\theta_{\mathcal{N}_i}=\{\theta_j:j\in\mathcal{N}_i  \}$ is set of neighboring parameters of agent $i$.  The objectives of optimal control is then to find the optimal control strategy $\theta_i^*, i\in(1,...,N)$ , such that  
\begin{equation}
	\begin{aligned}
		F(\{f_i(\theta_i^*, \theta_{\mathcal{N}_i}):i\in(1,...,N)\})\ge &\\F(\{f_i(\theta_i\neq \theta_i^*, \theta_{\mathcal{N}_i}):i\in(1,...,N)\}),
	\end{aligned}
\end{equation}
where $F(\cdot)$ denote the overall performance of all agents in the MAS. Such solution is often referred to as the Nash equilibrium strategy, meaning that can no further improvement can be made to individual solutions without deteriorating the performance. 
It has been found to be difficult to find a Nash equilibrium strategy while satisfying the system constraints \prettyref{eqn:systemConstraints}, and also limited by the nonlinear continuous system and the cost function with coupled parameters \cite{sefriouiNashGeneticAlgorithms2000, seung-mokleeCooperativeCoevolutionaryAlgorithmBased2015}. For the formation control problem which requires sufficient cooperation among neighboring agents, $F(\cdot)$ can be regarded as a simple summation of all individual $f_i(\cdot)$, and the formation control problem can be viewed as a constrained dynamic optimization problem with the objective of minimizing the total cost of formation error.  Thus the Nash equilibrium strategy can be obtained using a co-evolutionary algorithm that is designed for evolving simultaneous to reach the overall optimum fitness. In our previous work \cite{chenCooperativeGuidanceMultiple2022}, we improved on the natural evolutionary strategy (NES) and proposed an natural co-evolutionary strategy (NCES) algorithm that seeks global optimality for constrained multi-objective optimization problem in multi-agent system. In brief, NCES is a bio-inspired population-based algorithm that is capable of optimizing high-dimensional parameters, such as the weights of neural networks, towards the direction of higher fitness.   
The NCES algorithm usually proceeds as follows: first, the parameters $\theta_i$ for $i \in(1,...,N)$ are initialized, and the optimization objective is determined for a specific control problem. The fitness function $f(\cdot)$ which is embedded within the system model is thus obtained. Then, in the second step, iterative optimization is performed, and new perturbations $\epsilon_i $ are sampled $m$ times at the beginning of each iteration step which obey a probability distribution $p(\cdot)$ to obtain the perturbed population $\theta_i'=\theta_i+\epsilon_i$ corresponding to each agent, and their fitness values are then evaluated in the system in a distributed way, so that each population obtains the following corresponding gradient information
\begin{equation}
	g_{\theta_i}=\frac{1}{m\sigma^2}\sum_{i=1}^mf(\theta_i',\theta_{\mathcal{N}_i})\epsilon_i \prod_{c\in\mathcal{N}_i}p(\epsilon_c) ,
\end{equation}
and updates its parameters in a gradient ascent manner
\begin{equation}
	\theta_i=\theta_i + \eta_\alpha g_{\theta_i}.
\end{equation}
Finally, loop the second step until convergence or the Nash equilibrium strategy is found. For more detailed procedure of the algorithm, please refer to the original research.

\section{Applying Natural Co-evolutionary Strategy to formation control via neural networks}
\label{section:Implementation of Natural Co-evolutionary Strategy based formation control via neural networks}

\subsection{Distributed co-evolutionary strategy optimizing neural networks controller}
\begin{figure}[htbp]
 	\centering
 	\includegraphics[scale=0.15]{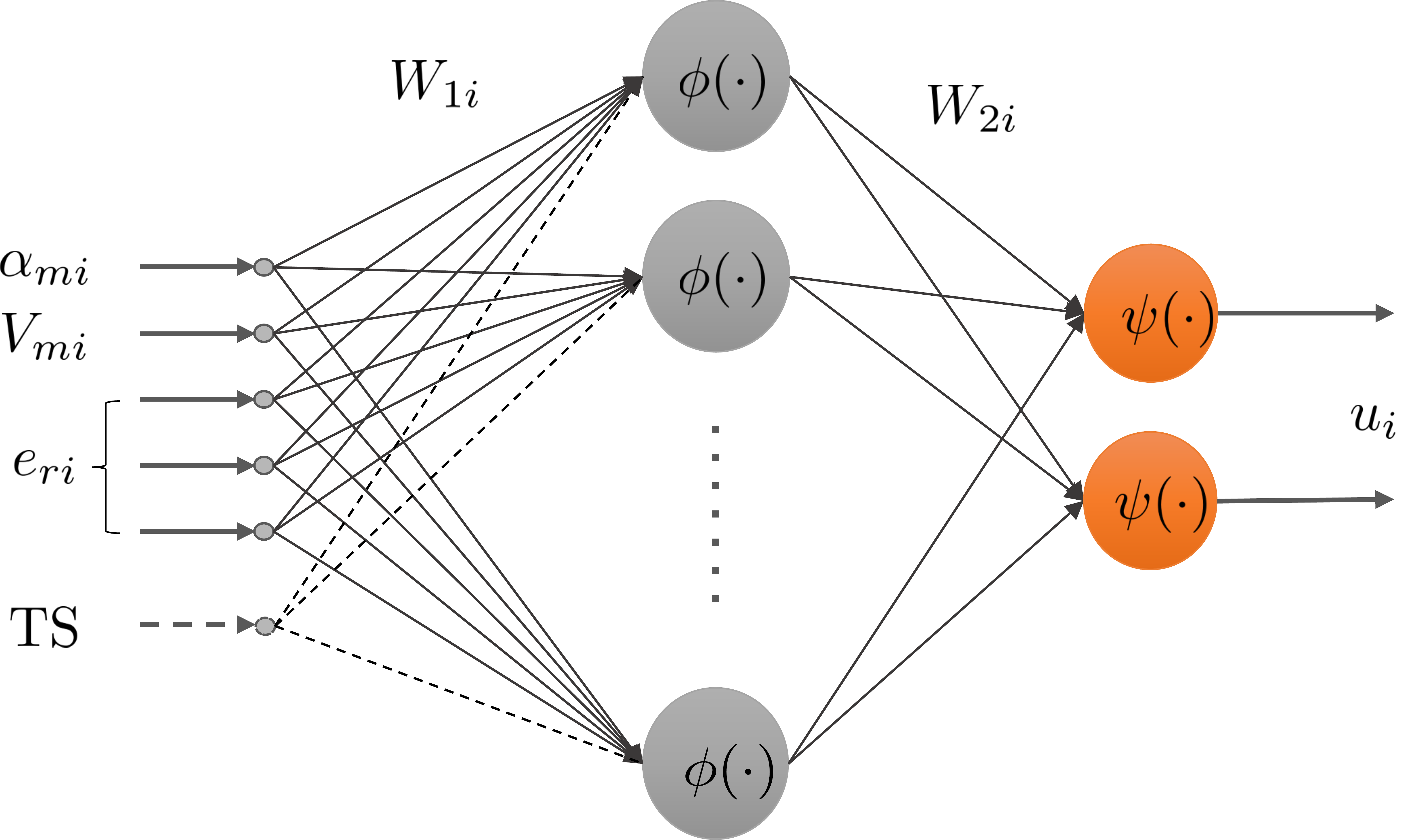}
 	\caption{Neural network controller schematic}
 	\label{fig:NeuralNetworkSchematic}
 \end{figure}

 Motivated by previous works which implement NNs in MAS, we use a multi-layer perceptron (MLP) NNs with a single hidden layer of 16 nodes as the distributed controller. The schematic of the neural networks controller is depicted in \prettyref{fig:NeuralNetworkSchematic}.

The weighting matrix for each layer are represented by $W_{1i} \  \text{and} \ W_{2i}$, with $\phi(\cdot)$ and $\psi(\cdot) $ as their activation function, respectively. Specifically, $\phi(\cdot)$ is the \emph{Sigmoid} function and $\psi(\cdot) $ is selected as the hyperbolic tangent function $(Tanh)$ in order to impose restrictions on the output to satisfy the system constraints. Another advantage of the neural network controller is shown that saturated control can be achieved by a reasonable choice of activation function without restricting the solution space. With $z=[\alpha_{mi}, V_{mi}, e_{ri}]^T$ denoting the input, the output of the neural network controller is 
\begin{equation}
	u_i = \psi(W_{2i}\cdot\phi(W_{1i}\cdot z)).
\end{equation}

Note that the dashed network input in the schematic represents the transform signal (TS), which is included in the input only when needed, as described in \prettyref{section:Simulation and result analysis}. The same network configuration will be used in the following experiments for all agents. Applying the previously mentioned NCES algorithm to train this controller, it is firstly determine that the parameters of the controller are $\theta_i = [W_{1i}, W_{2i}]$. The fitness function is the objective function \prettyref{eqn:objectiveFunction}, i.e., $f(\theta_i, \theta_{\mathcal{N}_i}) =J_i$ which is non-linearly coupled with the sates of neighboring agents and can be only evaluated through the interaction feedback within the system. In order to apply NCES to the training of the missile formation controller in this paper, the plain NCES algorithm is not sufficient to guarantee the global stability and the convergence speed of the algorithm. To further improve the performance of the algorithm and to perform specific optimization for the formation problem, we propose a series of supplementary techniques to effectively enhance the convergence speed and accuracy of the algorithm.

\subsection{Population adaptation technique}
In \cite{chenCooperativeGuidanceMultiple2022} we point out the importance of learning rate on the accuracy of the algorithm and proposed an algorithm for learning rate adaptation, however, adjusting the learning rate often consumes a lot of time, in order to ensure the speed of the algorithm here a novel population size adaptation algorithm is considered, which can be used to adjust the evolutionary process adaptatively.
 Based on the previous works \cite{nishidaPSACMAESCMAESPopulation2018, nomuraPrincipledLearningRate2022}, we can learn that the trend of the gradient is related to the complexity of the objective function, usually for the more complex optimization region such as multi-modal or noised function the estimated gradient is gentle, while in the flatter region such as the spherical space the gradient is relatively steep. To estimate the accuracy of the estimated natural gradient under the movement of the parameter distribution, the evolution path $\rho_{\theta}$ is introduced to detect the resistance in the evolutionary process. The population size is adapted based on the length of the evolutionary path following an empirical common sense that larger population size will lead to higher accuracy of the estimated gradient. Considering of the parameters of the $i^{th}$ agent as the vector consisting of the weighting matrices $W_{1i}, W_{2i}$, that is $\theta_i \in \mathbb{R}^{s}$, and $s$ is the total number of weights. The evolution path in iteration $t$ is calculated by accumulated the square of the Mahalanobis distance of the parameter movement of all agents as
\begin{equation}
	\rho_{\theta}(t)=\sum_{i=1}^{N}[\theta_i(t)-\theta_i(t-1)]^T\Sigma^{-1}[\theta_i(t)-\theta_i(t-1)],
\end{equation}
where $\Sigma$ is the covariance of the probability distribution from which the new populations are sampled, since the evolution path should not depend on the parameterization of the probability distribution. The population size $\eta_p(t)$ is then adjusted according to the evolutionary path as follows:
\begin{equation}
	\begin{aligned}
		\eta_{p}(t)&=\eta_p(t-1)(\beta+(1-\beta)\frac{\rho_{\theta}(t-1)}{\rho_{\theta}(t)}),\\
		\eta_p(t)&=\text{clip}( \max(\eta_p(t), \eta_p(t-1)), \eta_p^{\min}, \eta_p^{\max}),
	\end{aligned}
\end{equation}
where $\beta$ is a constant factor that determines the growth rate of the population size and $\eta_p^{\min}, \eta_p^{\max}$ are the minimum and maximum population size which are sent to the clip function $\text{clip}(\cdot)$in case of undesirable adapted values. Note that since the total optimization complexity tends to increase as evolution progresses, we adopt a non-decreasing strategy to adjust the population size to ensure stability. The initial population size is set to be $\eta_p(0) = 10 + 5\ln (s)$ referring to the default setup in \cite{glasmachersExponentialNaturalEvolution2010}, which should be a good candidate, and the boundaries are determined as
\begin{equation}
	 \eta_p^{\min}=\eta_p(0), \eta_p^{\max}=4\eta_p(0).
\end{equation}
Note that $s$ is the number of parameters of one agent, the above configuration is found to be appropriate according to experimental results. 
Through the empirical observation, we came to an intuitive conclusion that the increment of population size does not necessarily lead to the improvement of evolutionary quality, sometimes even leads to the difficulty of convergence or falls into the local optimum, which is presumably because a large population size will greatly average the contribution of individuals and thus reduce the exploratory nature of individual, so it is wise to adjust the size appropriately rather than just thinking that the larger the better.

\subsection{Cluster-based adaptive topology}
The inter-agent connectivity in the filed of multi-agent system have been primarily modeled and characterized by means of graph theory \cite{desaiControllingFormationsMultiple1998, ohSurveyMultiagentFormation2015}, which in this paper will be utilized to identify the observable neighbors of the missiles in the formation. In this subsection we review some of the basics of graph theory. 
Suppose there are $n$ agents in the MAS (agents can be represented by nodes), and a graphs $\mathcal{G}$ is defined as $\mathcal{(\mathcal{V}, \epsilon)}$, where $\mathcal{V}=(v_1,v_2,...,v_n)$ represents the set of nodes, and $\epsilon \subseteq\mathcal{V}\times\mathcal{V}$ represents the set of edges composed of directed connections of different nodes. The neighbor set of node $i$ is defined as $\mathcal{N}_i=\{j\in\mathcal{V}:(i,j)\in\epsilon\}$.  $j\in \mathcal{V}$ is said to be connected to $i\in\mathcal{V}$ in a directed way if $(i,j)\in \epsilon$ and $(j,i)\notin \epsilon$, otherwise $j $ is said to be connected to $i$ in an undirected way if both $(i,j)$ and $(j.i)$ belongs to $\epsilon$ . A directed path of $\mathcal{G}$ is a series of adjacent edges of the form$(v_1, v_2),(v_2,v_3),...,(v_i, v_j),(v_j,v_k)$, and a graph is said to have a spanning tree if there exist a directed path from one node to any other nodes, and a graph is said to be connected in directed (undirected) manners if there is a directed (undirected) path between any pair of distinct nodes.  We use the adjacency matrix $A=[a_{ij}]\in \mathbb{R}^{|\mathcal{V}|\times|\mathcal{V}|}$ to represent the above node connectivity, in which
\begin{equation}
	a_{ij}=\begin{cases} 
		0, &\text{if i=j or $(i,j)\notin\epsilon$}\\
		1, &\text{otherwise}
	\end{cases}.
\end{equation}
The associated Laplacian matrix is $L=D-A$. $D=diag(a_i)$ is the diagonal matrix of vertex degrees where $a_i$ is the degree of vertex $i$,  and $a_i=\sum_{j=1}^na_{ij}$. The Laplacian matrix $L$ is symmetric semi-positive definite matrix.

One of the foremost goal of designing communication method is to ensure highly reliable and low latency communication for all nodes in the swarm. The communication quality of missile swarm is easily affected by various conditions especially when performing in hostile environment.
In another word, swarm of missiles are constantly subject to threats from enemy’s air defense system during flight, and some missiles may get intercepted and lost communication with the rest of the swam. Therefore, dynamic communication approaches needs to ensure that when some nodes fail, the swarm can still maintain effective communication to complete the mission.
Based on the previous work in network topology and swarm communication, we developed a novel adaptive cluster-based network, which adaptively reconfigure the communication topology to achieve robust and fault-tolerant communication.
Under the guidance of the well known Molloy-Reed criterion \cite{barabasiNETWORKSCIENCENETWORK}, which is
\begin{equation}
	k=\frac{<k^2>}{<k>}>2,
\end{equation}
$k$ denotes the average degree of an arbitrary node which can be considered somehow as the node degrees $a_i$, it can be inferred that each node in the communication network should be at least connected to two other nodes. What is also known is that for displacement-based formation, the formation is persistent only if there exists at least a spinning-tree in the communication topology. When establishing a communication connection, the formation error tends to grow as the communication distance between nodes increases, so we favour the connection method which minimizes the length of the communication chains \cite{dasVisionbasedFormationControl2002}.
With the minimal constraint of satisfying the above conditions, we propose to set one node(usually node $1$) of the network as the cluster head and the other nodes choose another communication node according to the inter-node distances that are derived from the formation definition. For achieving this, we define the inter-agent distance matrix as 
\begin{equation}
	D_g = [d_{ij}]\in \R^{n\times n}, \ d_{ij}=||\lambda_{pi}-\lambda_{pj}||,
\end{equation}
where $d_{ij} $ is the Euclidean distance between nodes that can be calculated from the definition of the formation in \prettyref{section:Preliminaries and Problem Formulation}, and $D_g$ is symmetric positive definite matrix. Given that node $h$ is selected as the cluster head, the element of the adjacency matrix is determined as 
\begin{equation}
	\begin{aligned}
		a_{ij} = \begin{cases}
			1, &\text{if $i\ne h$, and $j\in$}\{\min\{k=\argmin_k{d_{ik}}:\\&k\in{\N},1\le k\le n, k\ne h\},h\};\\
			0, &\text{otherwise;}
		\end{cases}.
	\end{aligned}
\end{equation}

In the proposed adaptive topology, the cluster head undertakes the task of broadcasting its own state to other nodes in the network, or follower nodes need to follow the cluster head, in either case the cluster head is not constrained by nodes other than the reference target. It is also important to note that only the cluster head has access to the reference target information, which greatly reduces the communication or detection burden of follower nodes.
\begin{figure}[htbp]
	\centering
	\begin{subfigure}[h]{0.45\columnwidth}
		\centering
		\includegraphics[scale=0.5]{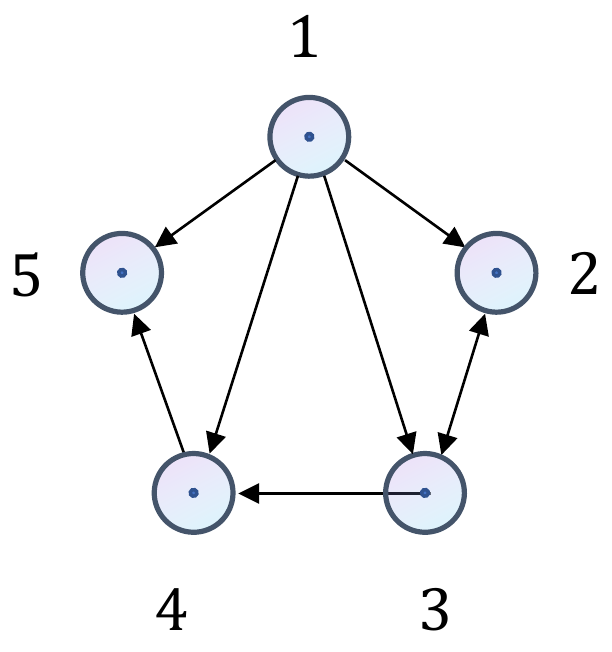}
		\caption{all nodes are valid}
		\label{fig:adaptiveTopology_fiveNodes}
	\end{subfigure}
	\begin{subfigure}[h]{0.45\columnwidth}
		\centering
		\includegraphics[scale=0.5]{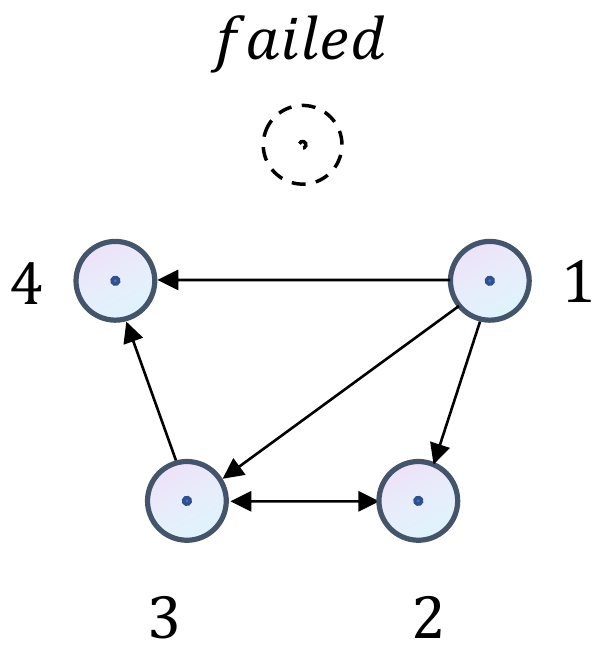}
		\caption{node 1 is failed}
		\label{fig:adaptiveTopology_failedNodes}
	\end{subfigure}
	\caption{Schematic diagram of the adaptive topology network with 5 nodes}
\end{figure}

Using the above communication configuration method for a network with 5 nodes as an example, the obtained communication topology is shown in \prettyref{fig:adaptiveTopology_fiveNodes} and its adjacent matrix is 
\begin{equation}
	A^1=\begin{bmatrix}
		0&0&0&0&0\\
		1&0&1&0&0\\
		1&1&0&0&0\\
		1&0&1&0&0\\
		1&0&0&1&0
	\end{bmatrix}
\end{equation}
Node 1 is the cluster head by default. When node 1 fails, node 2 successively inherits the head position as shown in \prettyref{fig:adaptiveTopology_failedNodes}, and the adjacent matrix becomes
\begin{equation}
	A^2=\begin{bmatrix}
		0&0&0&0\\
		1&0&1&0\\
		1&1&0&0\\
		1&0&1&0
	\end{bmatrix}
\end{equation}

Using the proposed adaptive communication method, the network connectedness can be ensured no matter how many nodes fail. The proof is that graph $\mathcal{G}$ is uniformly connected since for any $t\ge t_0$, there exists a node $h\in \mathcal{V}$ such that $h$ is the root of a spanning tree \cite{linStateAgreementContinuous2007}. 

\subsection{Model-based constrained policy}
The NCES algorithm explores the optimal policy that maximizes fitness by continuously interacting with the environment; consequently, the NCES algorithm and the reinforcement learning algorithm are prone to fail to achieve global convergence of the control policy in many cases due to excessive cumulative time or inherent defects in the design of the objective function. In much of the previous literature, control policies were allowed to freely explore regions of the environment under system constraints. In recent years, the idea of constrained policy has gradually emerged \cite{achiamConstrainedPolicyOptimization2017}, which states that control policies should be executed under constraints that can be imposed either by human-developed rules or by the feedback state of the system. Constrained policy is also known as safety exploration \cite{rayBenchmarkingSafeExploration} has been increasingly applied to simulation and training of realistic robotic controllers, by which not only can personal safety be ensured during the training process, but also global convergence can be accelerated to some extent.
Based on the above facts, we propose a model-based constrained policy strategy. The non-linear error dynamic of the second-order system has been obtained in \prettyref{eqn:errorDynamic}, which is a function of the system input. To apply the method, we assume that the control input of the communication object can be obtained by the agent, so that at time $t$, the predicted formation error at next step can be calculated as follows:
\begin{equation}
	\hat{e}_{ri}(t) = e_{ri}(t)+\dot{e}_{ri}(t)\tau,
\end{equation}
where $\tau$ is the time step, and the error deviation $\triangle e_{ri}(t)\in\R^3$ is 
\begin{equation}
	\triangle e_{ri}(t)= |\hat{e}_{ri}(t)|-|e_{ri}(t)|,
\end{equation}
then the aggregate matrix is defined as $\triangle E(t)=[...\triangle e_{ri}(t)...]\in\R^{3\times n}$, for $i\in \{1,2,...,n\}$ .  A termination indicator $ST$ is assigned to the system so that if it is a nonzero value, the system terminates and restart the training algorithm, and it is defined as 
\begin{equation}
	ST(t)= \begin{cases}
		1, &\text{if $\text{min}(\triangle E(t))\gt\delta_s$;}\\
		0, &\text{otherwise,}
	\end{cases}
\end{equation}
 and $\delta_s$ is the threshold that measures the maximum error increment that the algorithm can tolerate. It satisfies $\delta_s \lt (2l+1)\tau V_{max}$, where $l $ is the number of neighbors for the agent with the most neighbors in the MAS. With a reasonable choice of $\delta_s$, the algorithm is expected to exclude control strategies that deviate too much from the desired trajectory at an early stage. Experimental evidence shows that without such a constrained policy, the convergence speed and global optimality of the algorithm will be degraded.
Applying the above adaptation techniques, the detailed pseudo-code of the proposed  algorithm that is used to train the missile formation controllers of interest is shown in \prettyref{alg:fcalgorithm}. Since the network topology adaptation and constrained policy strategies are embedded in the process of fitness evaluation, they are not exhibited in the pseudo-code.
\begin{algorithm}
	\caption{The distributed NCES based formation control algorithm}
	\label{alg:fcalgorithm}
	\begin{algorithmic}
	 \Require agent number $N\in\mathbb{N}^+$, population size $\eta_p$, standard deviation $\sigma$, $\beta$, evolution path $\rho_{\theta}$, number of parameters $m$, iteration $t$
	\Initialize
	\For{ each agent $i\in\{1,...,N\}$}
	\State initialize parameter $\theta_i^{init}$   
	\State $t\leftarrow 0$
	\State $\theta_i(0) \leftarrow \theta_i^{init}$
	\State $\eta_p(0) \leftarrow 10 + 5\ln (m)$
	\State $\eta_p^{min} \leftarrow \eta_p(0)$
	\State $\eta_p^{max} \leftarrow 4\eta_p(0)$
	\EndFor
	\EndInitialize
	\While{stopping criterion not met}
	\State $t\leftarrow t+1$
	\For{k = 1,...,m}
	\For{each agent $i\in\{1,...,N\} $ }
	\State sample $\epsilon_i^k\sim N(0,\sigma^2 I)$
	\State $\theta_i^k\leftarrow\theta_i+\epsilon_i^k$
	\EndFor
	\State evaluate fitness $f(\theta_i^k, \theta_{\mathcal{N_i}}^k),$ for $i\in\{1,...,N\}$
	\EndFor
	
	\For{each agent $i\in\{1,...,N\} $ }
	\State calculate natural gradient:
	\State $g_{\theta_i}\leftarrow\frac{1}{m\sigma^2}\sum_{k=1}^{m}f(\theta_i^k,\theta_{\mathcal{N_i}}^k)\epsilon_i^k \prod_{c\in\mathcal{N}_i}p(\epsilon_c^k)$
	\State $\theta_i(t)\leftarrow \theta_i(t-1)+\ \eta_\alpha\cdot g_{\theta_i}$
	\EndFor
	\State append evolution path: \State $\rho_{\theta}(t)\leftarrow\sum_{i=1}^{N}[\theta_i(t)-\theta_i(t-1)]^T\Sigma^{-1}[\theta_i(t)-\theta_i(t-1)]$
	\State $\text{AdaptPopulationSize($\rho_{\theta}$)}$
	\EndWhile
	\end{algorithmic}
\end{algorithm}
\begin{algorithm}
	\caption{AdaptPopulationSize($\rho_{\theta}$)}
	\label{alg:adaptPopSize}
	\begin{algorithmic}
		\If{$\text{length($\rho_{\theta}$)}\gt1$}
		\State  $\eta_{p}(t)\leftarrow\eta_p(t-1)(\beta+(1-\beta)\frac{\rho_{\theta}(t-1)}{\rho_{\theta}(t)})$
		\State  $\eta_p(t)\leftarrow\text{clip}( \max(\eta_p(t), \eta_p(t-1)), \eta_p^{\min}, \eta_p^{\max})$
		\Else
		\State  $\eta_p(t)\leftarrow \eta_p(0)$
		\EndIf
	\end{algorithmic}
\end{algorithm}

\section{Simulation and result analysis}
\label{section:Simulation and result analysis}

In this section various experiments are implemented to demonstrate the effectiveness of the proposed formation control algorithm. In order to simulate actual physical environment, the construction of the experimental scenario and the modeling of the multi-agent system were carried out in PyBullet \cite{coumans2016pybullet}. We simulated the operation of the missile swarm in three-dimensional space at a fixed altitude, and for convenience the trajectories was plotted as two-dimensional planar graphs. Some complex aerodynamic parameters such as air resistance are removed, while collision detection is preserved. The simulation timestep $\tau$ is set to $0.1$ for the following experiments.

\subsection{Basic formation control}
 First of all the basic formation control tasks  are implement to examine the validity of the algorithm under basic situations, the missile swarm of five missiles entails tracking the reference target moving along diagonal or spiral trajectory while keeping the formation geometric. The reference target, or virtual leader is set to be at the center of the formation in order to achieve error-free formation control. The objective is to achieve zero tracking error as well as zero formation maintenance error for all time, that is $\int_t |e_{ri}(t)|\rm{d}t \to 0_{3}, $ for $i \in \{1,...,N\}$. The hyper-parameters of the algorithm is listed in \prettyref{tab:hyperParameters} and it is noted that this parameter setting applies to all of the following experiments. Additionally, the system constraints are shown in \prettyref{tab:systemConstraints}, the missiles as well as the reference target are subjected to saturated control input and limited states. For the linear trajectory, the target is driven by constant control input, while for the spiral trajectory, the control input of the target is $v_r=0.65-0.01t \text{(km/s)}$ and $\omega_r = 0.1+0.01t\text{(rad/s)}$. 
 
 \begin{table}[]
 	\centering
 	\caption{Hyper-parameters of the proposed algorithm}
 	\label{tab:hyperParameters}
 	\begin{tabular}{@{}lp{0.5\columnwidth}l@{}}
 		\hline
 		Symbol        & Description                       & Value               \\ \hline
 		$\eta_\alpha$ & learning rate                     & 0.02                \\
 		$\tau$        & time step                         & 0.1                 \\
 		$\sigma$      & standard deviation                & 0.2                 \\
 		$\beta$       & population size adaptation factor & 0.84                \\
 		$K_{c}$       & cost weight matrix                & $[0.15,0.15,0.1]^T$ \\ \hline
 	\end{tabular}
 \end{table}
\begin{table}[]
	\centering
	\caption{System Constraints}
	\label{tab:systemConstraints}
	\begin{tabular}{@{}lp{0.5\columnwidth}l@{}}
		\hline
		Symbol     & Description                                           & Value    \\ \hline
		$V_{max}$  & maximum velocity of both missile and reference target & 0.8 km/s \\
		$V_{min}$  & minimum velocity of both missile and reference target & 0.3 km/s \\
		$a_{lmax}$ & maximum lateral acceleration                          & 40g      \\
		$a_{vmax}$ & maximum velocity acceleration                         & 30g      \\ \hline
	\end{tabular}
\end{table}
 
\begin{figure}[htbp]
 	\centering
 	\includegraphics[scale=0.4]{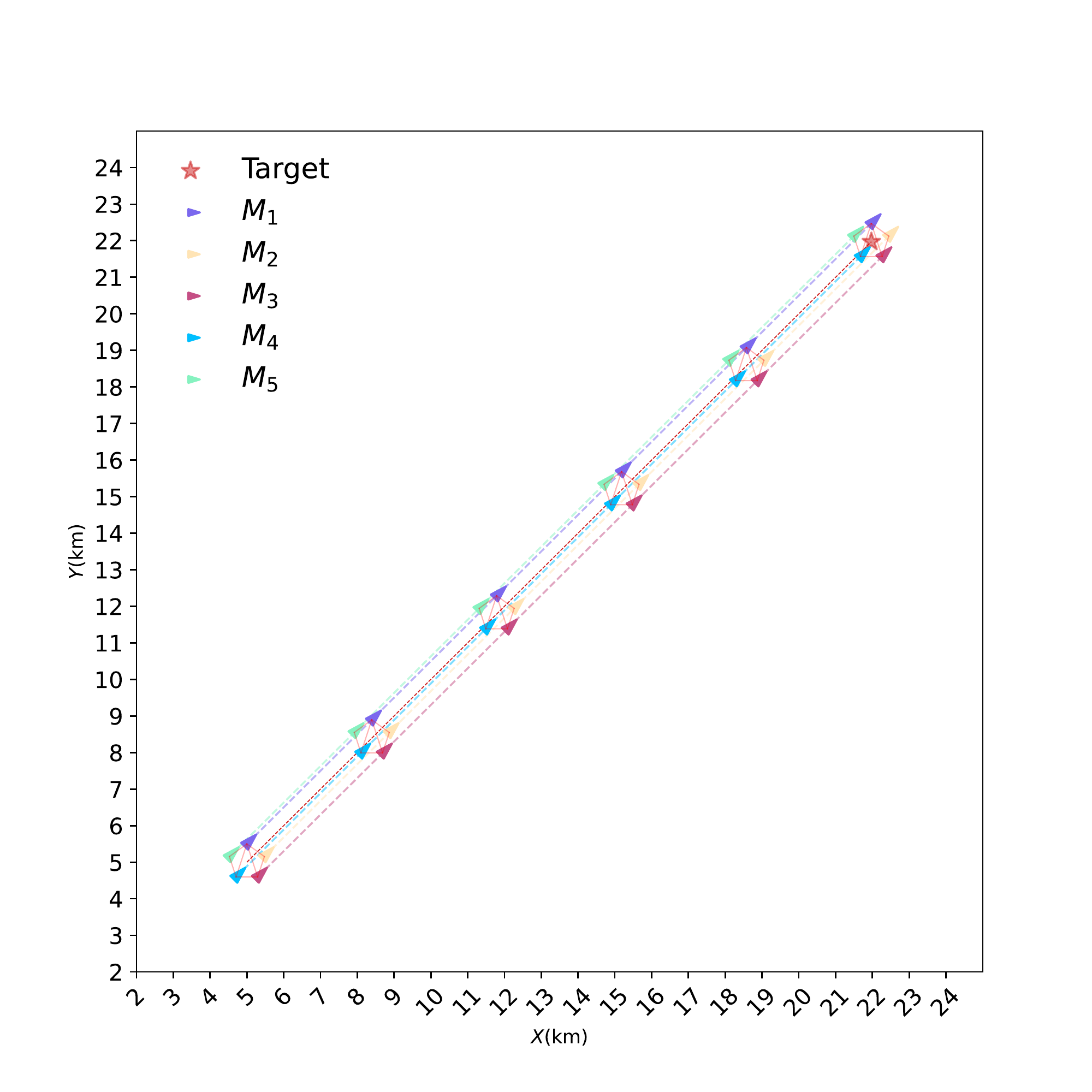}
 	\caption{Trajectories of basic formation control along linear trajectory}
 	\label{fig:BasiclinearTraj}
 \end{figure}

\begin{figure}[htbp]
 	\centering
 	\includegraphics[scale=0.4]{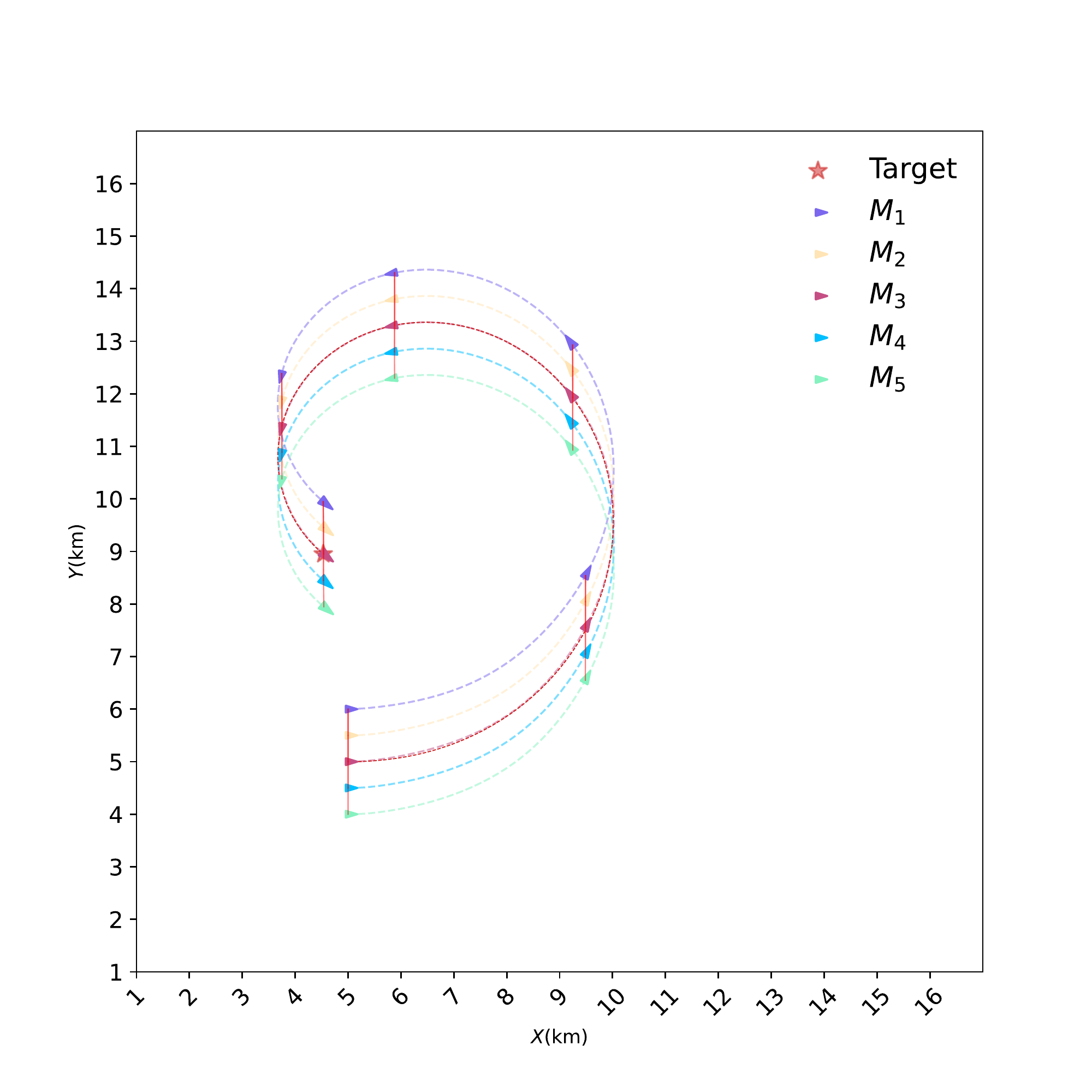}
 	\caption{Trajectories of basic formation control along spiral trajectory}
 	\label{fig:BasicspiralTraj}
 \end{figure}

\begin{figure}[h]
	\centering
	\begin{subfigure}[ht]{1\columnwidth}
		\centering
		\includegraphics[scale=0.5]{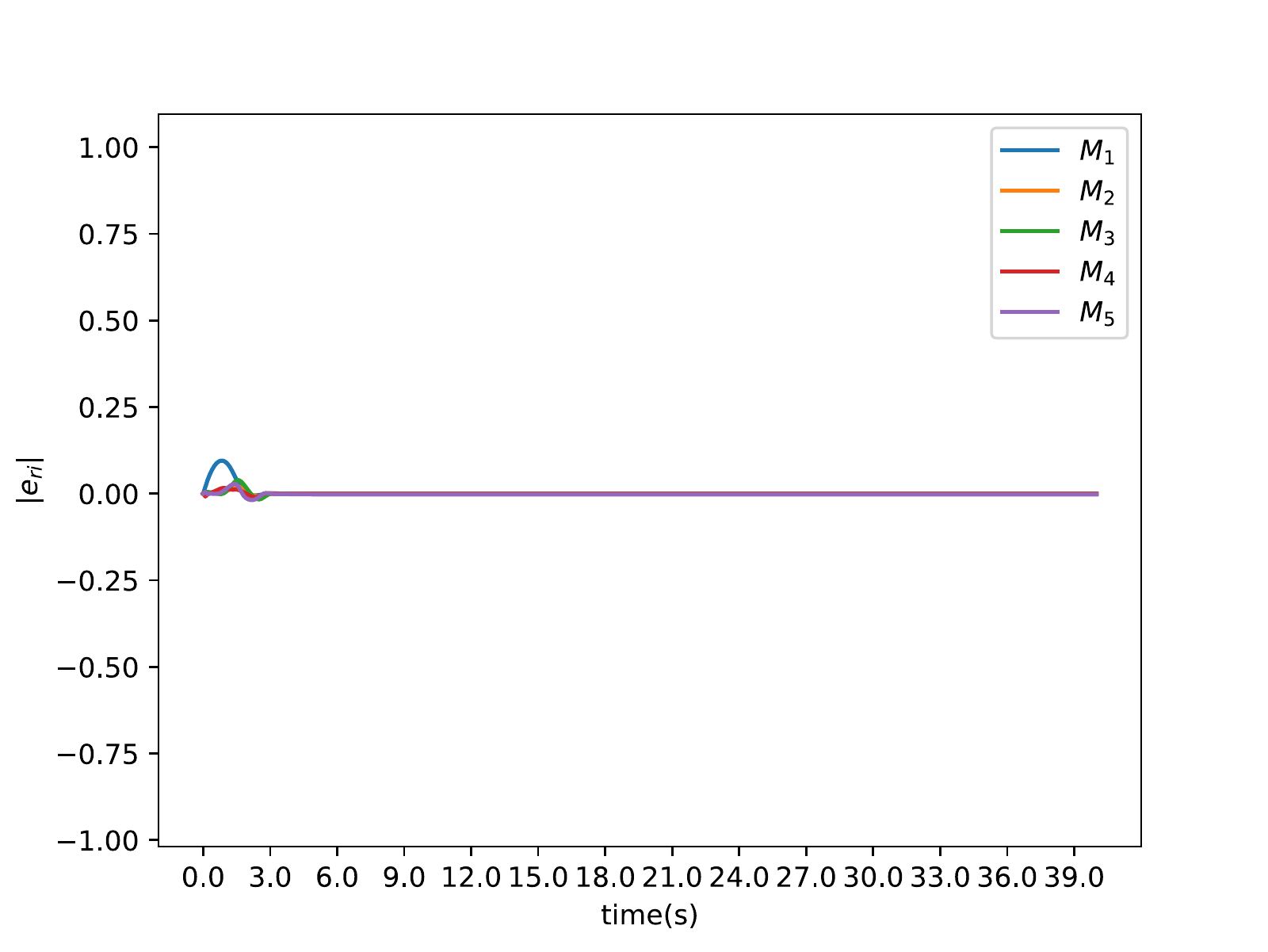}
		\caption{resultant error curves}
		\label{fig:BALinearResError}
	\end{subfigure}
	
	\begin{subfigure}[ht]{1\columnwidth}
		\centering
		\includegraphics[scale=0.5]{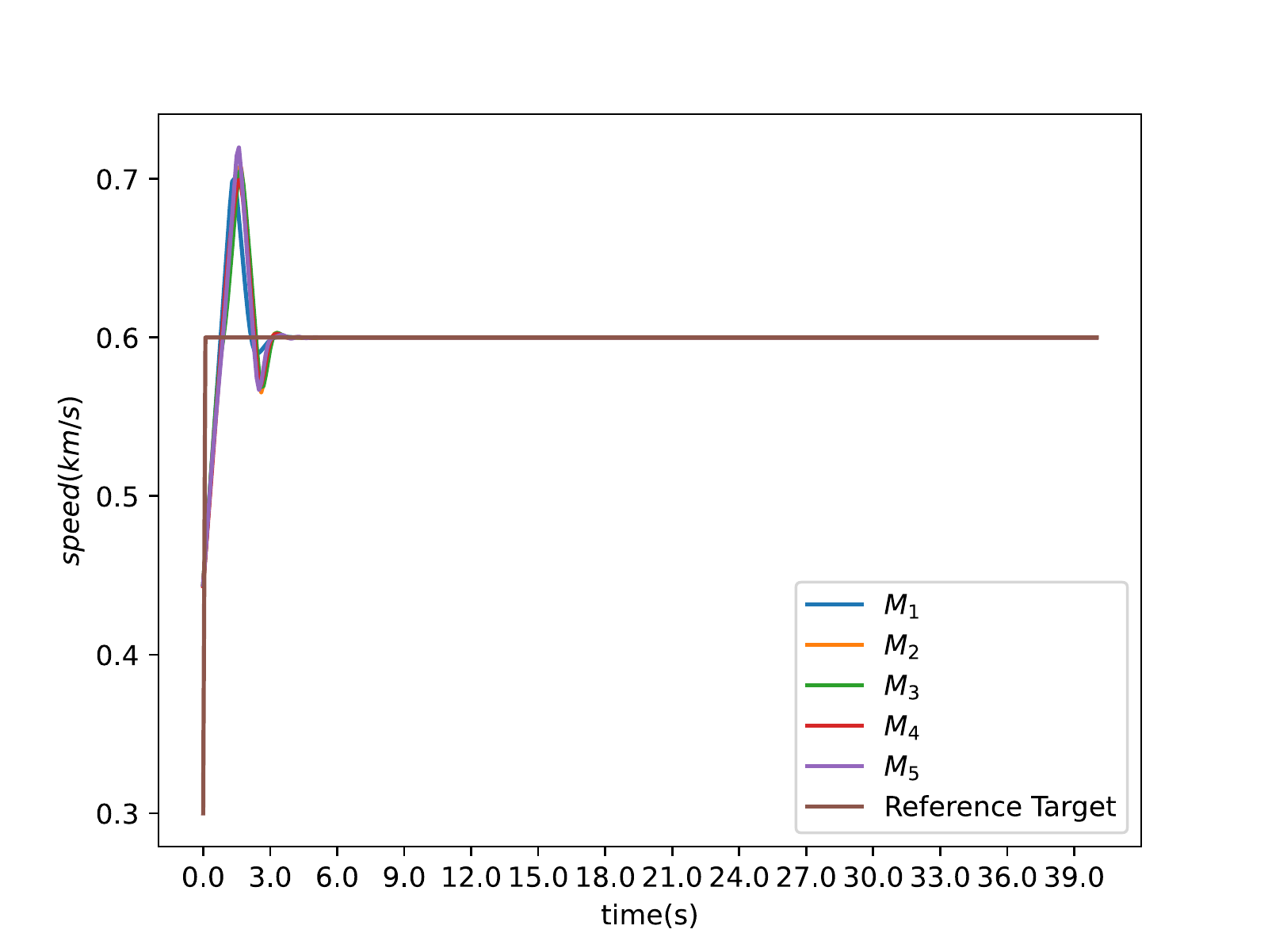}
		\caption{speed curves}
		\label{fig:BALinearSpeed}
	\end{subfigure}

	\begin{subfigure}[ht]{1\columnwidth}
		\centering
		\includegraphics[scale=0.5]{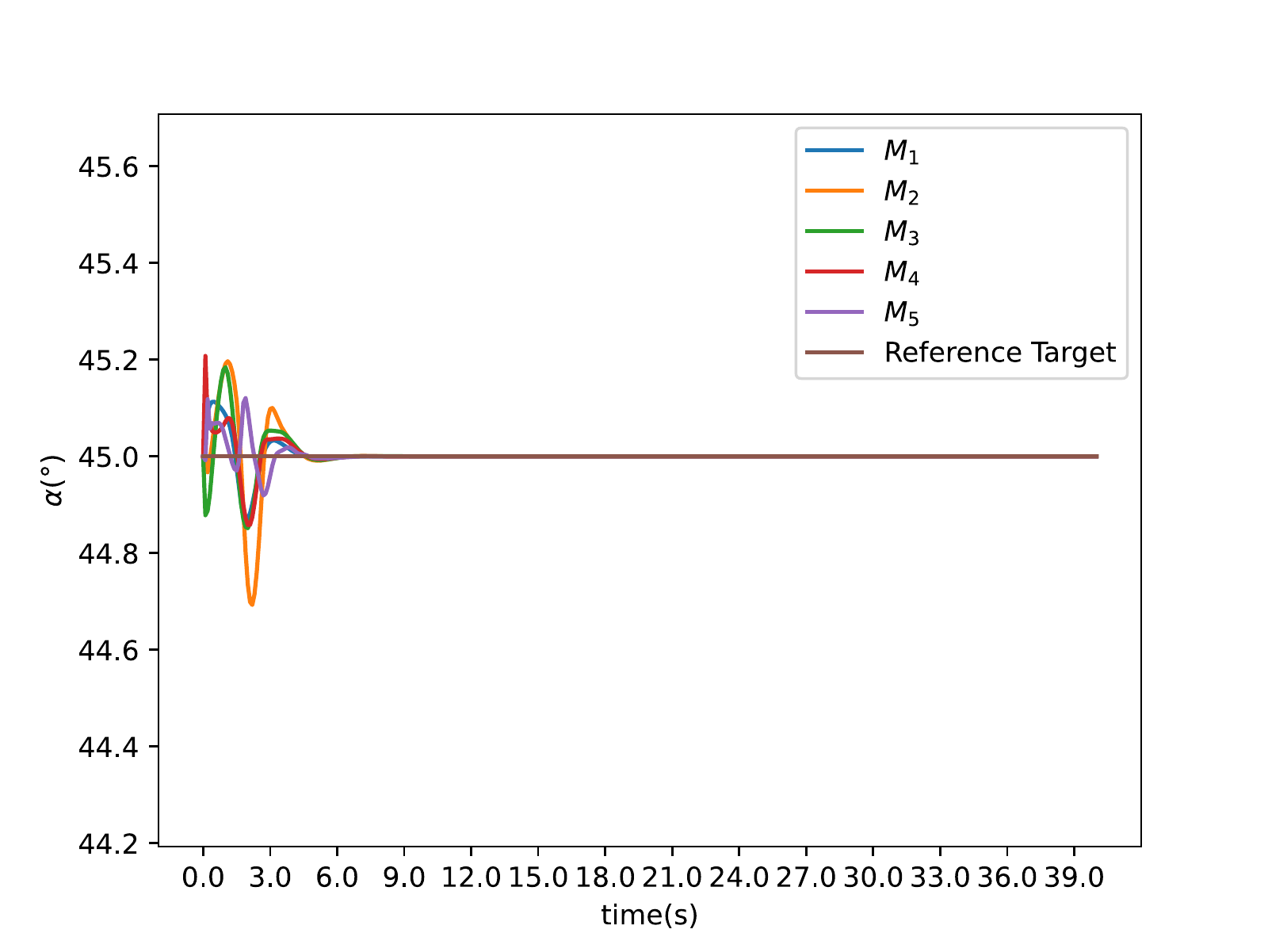}
		\caption{heading angle curves}
		\label{fig:BALinearAngle}
	\end{subfigure}
	\caption{Analytical results of the linear trajectory case}\label{fig:BasicAnalyticLinear}
\end{figure}

\begin{figure}[h]
 	\centering
 	\begin{subfigure}[ht]{1\columnwidth}
 		\centering
 		\includegraphics[scale=0.5]{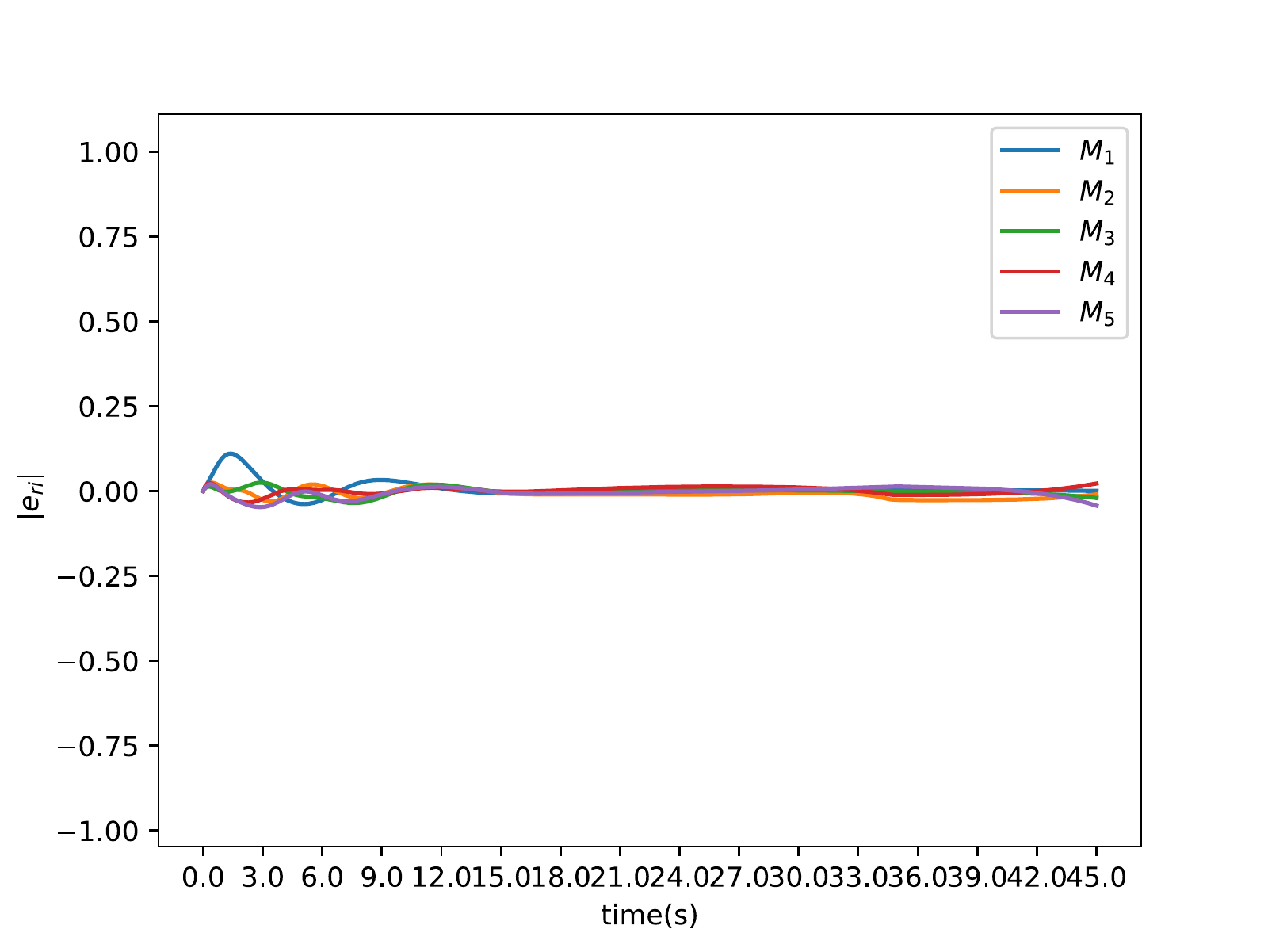}
 		\caption{resultant error curves}
 		\label{fig:BASpiralResError}
 	\end{subfigure}
 	
 	\begin{subfigure}[ht]{1\columnwidth}
 		\centering
 		\includegraphics[scale=0.5]{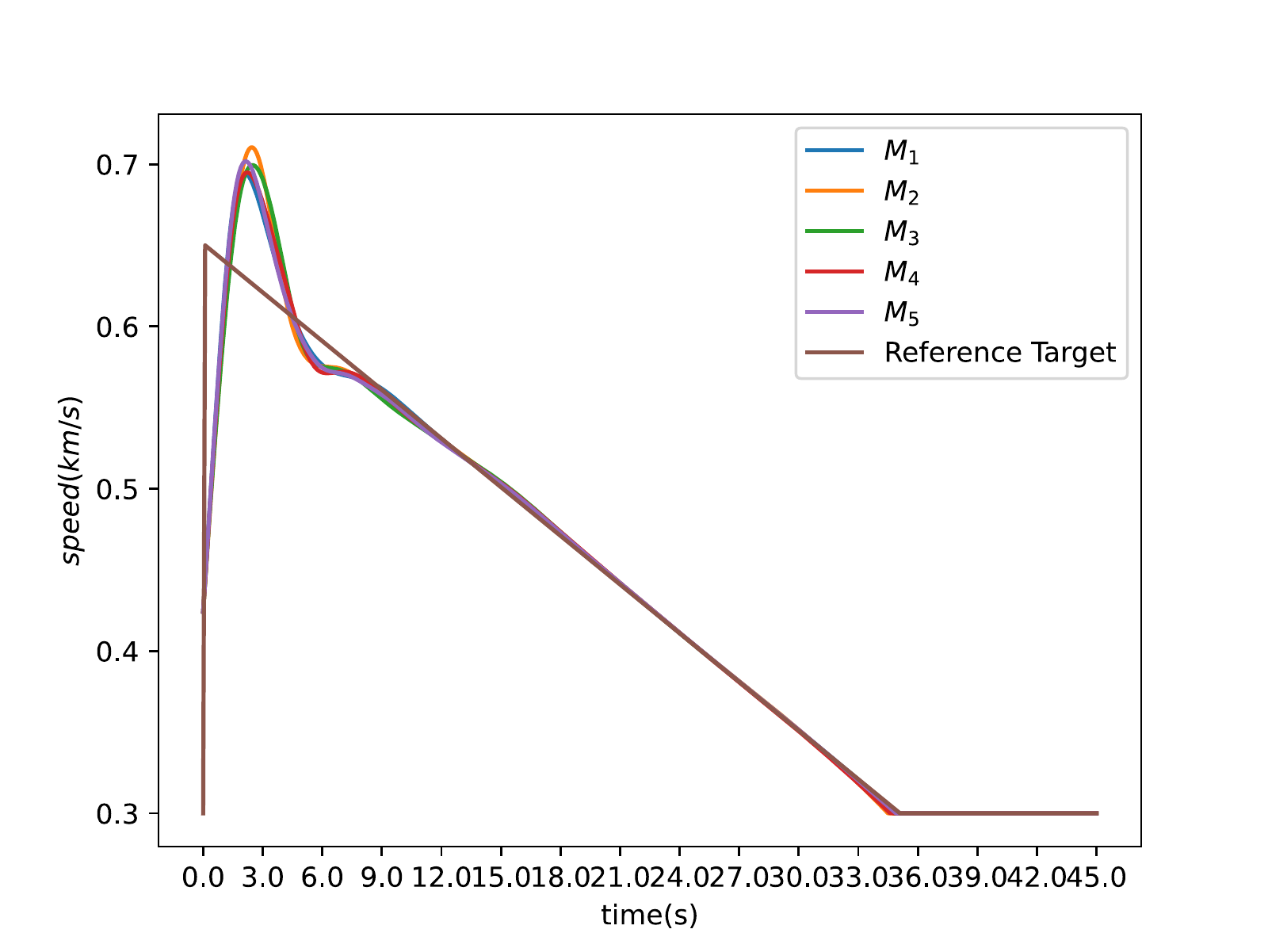}
 		\caption{speed curves}
 		\label{fig:BASpiralSpeed}
 	\end{subfigure}
 	\begin{subfigure}[ht]{1\columnwidth}
 		\centering
 		\includegraphics[scale=0.5]{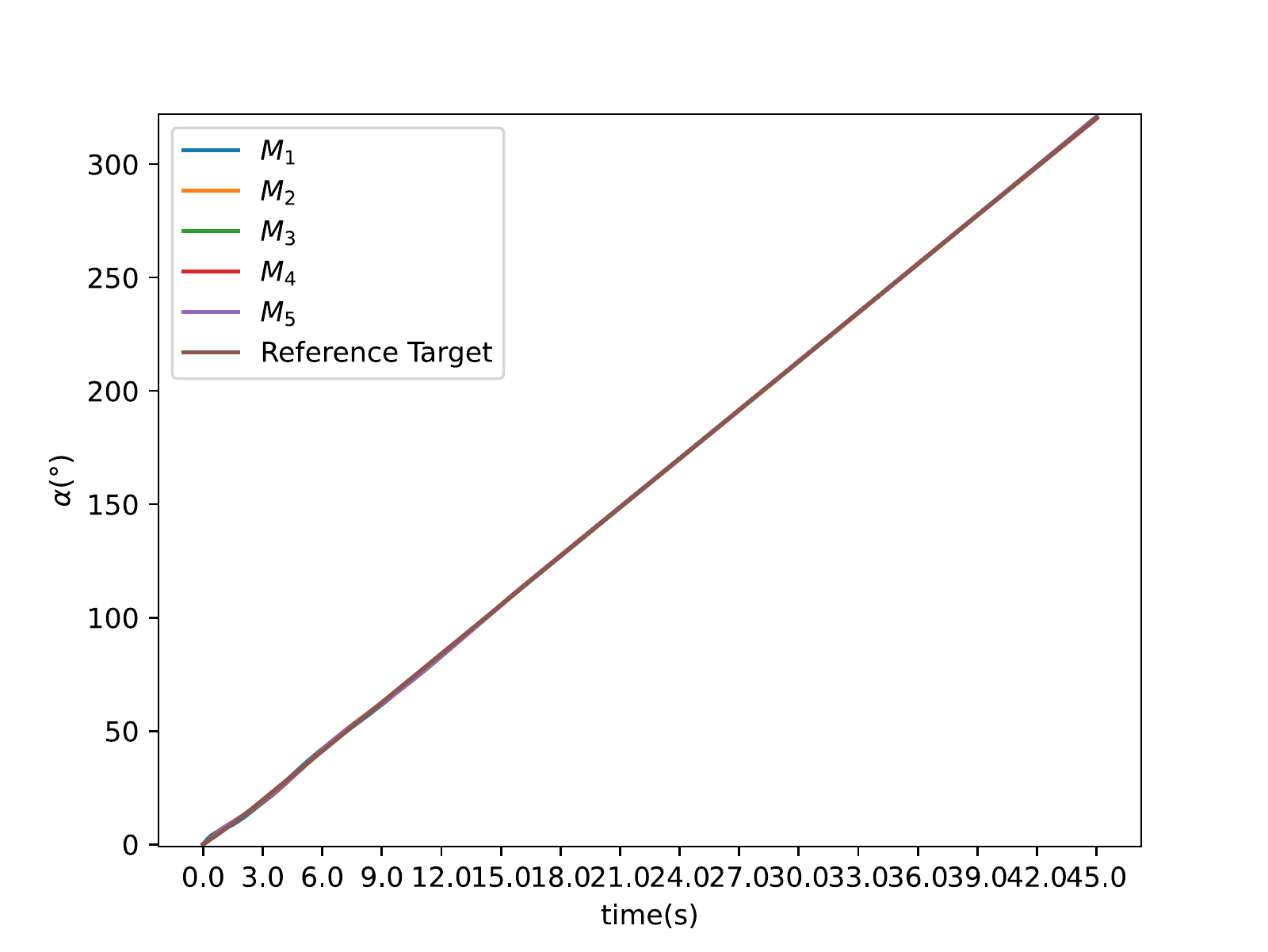}
 		\caption{heading angle curves }
 		\label{fig:BASpiralAngle}
 	\end{subfigure}
 	\caption{Analytical results of the spiral trajectory case}\label{fig:BasicAnalyticSpiral}
 \end{figure}

 For convenience, we use $|e_{ri}|$ to represent the resultant error of each agent. The trajectories of the two situations are shown in \prettyref{fig:BasiclinearTraj} and \prettyref{fig:BasicspiralTraj}, and the corresponding analytical results are presented in \prettyref{fig:BasicAnalyticLinear} and \prettyref{fig:BasicAnalyticSpiral}. It can be observed that the resultant error is maintained in a small interval (within 0.1) in both cases and that a good synchronization of the velocity and heading angles among the missiles is achieved, despite the fact that velocity information is not provided.

 
\subsection{Moving into formation}

\begin{figure}[htbp]
	\centering
	\includegraphics[scale=0.4]{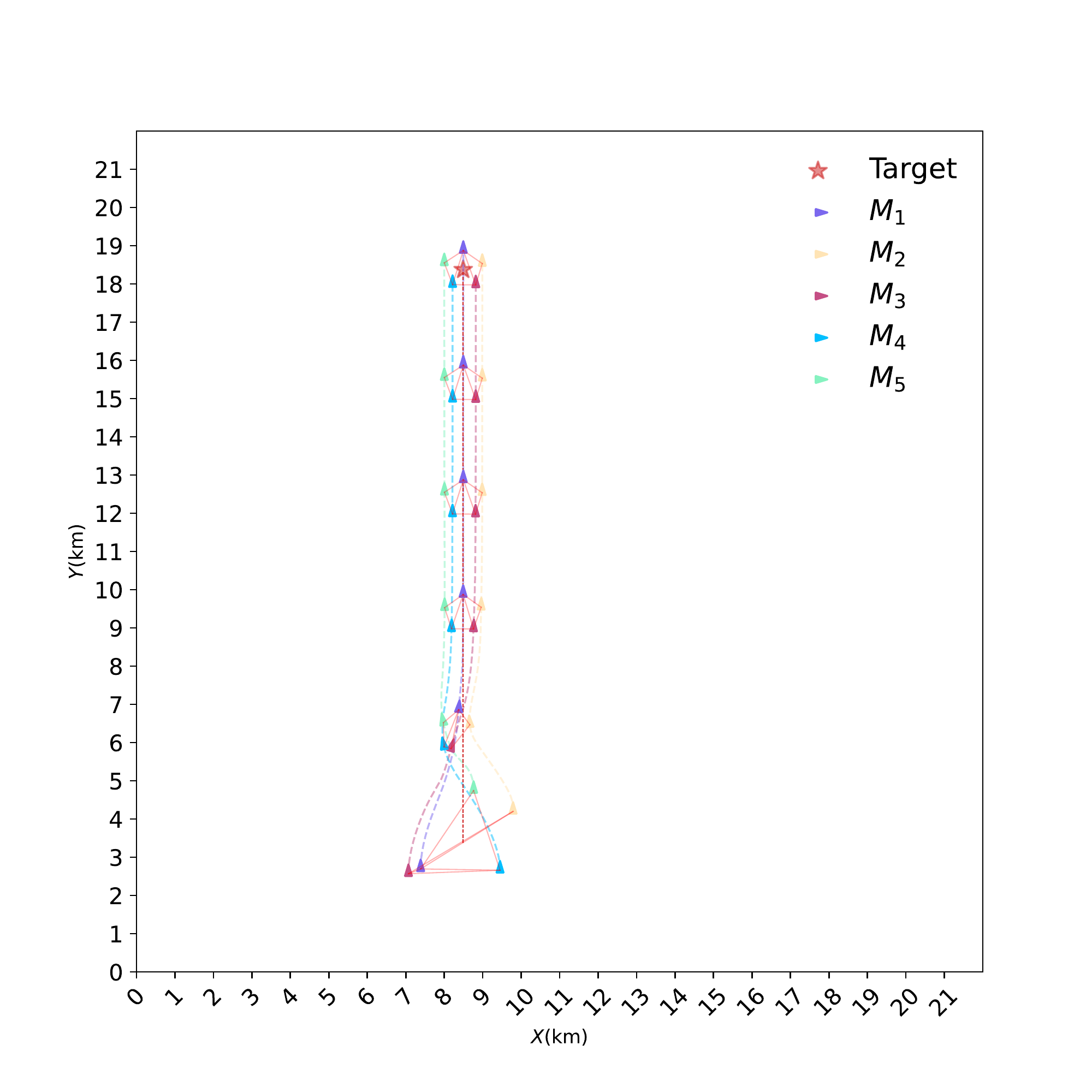}
	\caption{Trajectories of moving into formation case}
	\label{fig:mIfTraj}
\end{figure}

\begin{figure}[htbp]
	\centering
	\includegraphics[scale=0.5]{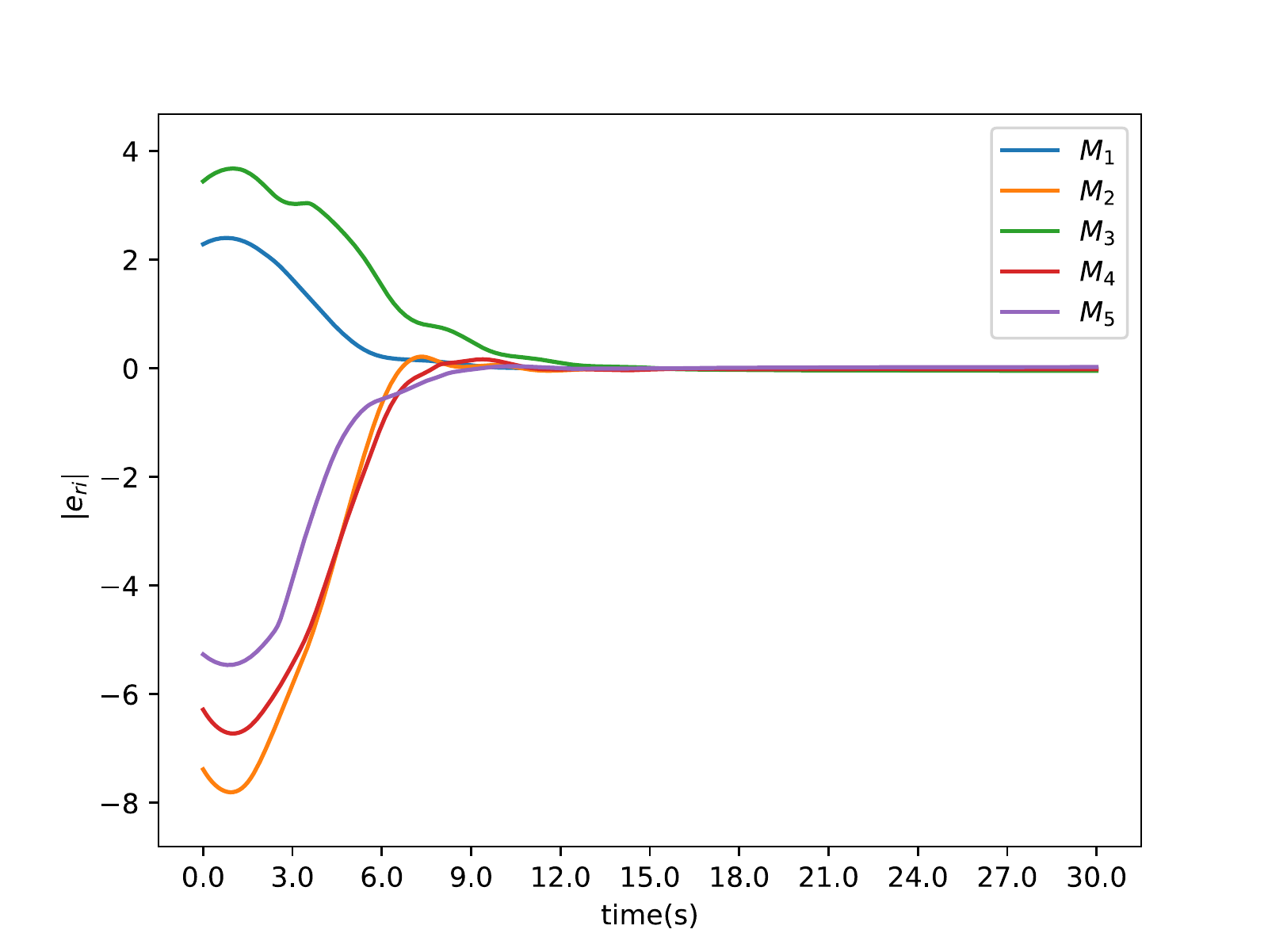}
	\caption{Resultant error curves of moving into formation case}
	\label{fig:resErrorMvForm}
\end{figure}

Moving into formation is however different from cases where the formation is in the ideal geometry in the initial state. The missiles are separated from the reference target with their positions initialized randomly in an area of 4 km wide and 3 km long. The missiles need to first move to the designated formation shape and then track the reference target in a consistent motion direction, and in this case, the target moves along the y-axis with a constant speed of 0.5km. \prettyref{fig:mIfTraj} shows the motion trajectory of the swarm formation, and \prettyref{fig:resErrorMvForm} shows the resultant error during the flight. It is evident from the results that the formation is able to adjust to the desired formation shape rapidly (mostly within 10 seconds) and achieve high accuracy in tracking and maintaining the formation in the case of random initial distribution. 

\subsection{Switching formations}

Further, we discuss the case in which the missile swarm are supposed to switch among formation patterns in order to avoid obstacles and go through narrow spaces. To achieve this transformation, an additional node $TS$ is appended to the input layer of the policy network as depicted in \prettyref{fig:NeuralNetworkSchematic}. It’s assumed that whenever the obstacle is detected by the leader missile, or more literally the cluster head, the missile then send a signal to all connected missiles to perform formation switch $\lambda^R\to \lambda^L$. Although formation geometries are predefined by the controller, the shapes can be controlled by the parameter $l_f$ and $\beta$, which is changeable during the flight.

In this scenario, two walls are placed in the trajectory of the formation flight to create narrow spaces, and the missiles need to negotiate the obstacles by changing the shape or size of the formation. We implement a event-based formation switching strategy, in which nodes in the formation are able to detect obstacles within a certain range $d_c$ and send a formation switching signal to all other nodes if an obstacle is detected. Similarly, after crossing the obstacle and reaching a safe distance, a formation recovery signal will be sent to restore the original formation.

\begin{figure}[htbp]
	\centering
	\includegraphics[scale=0.4]{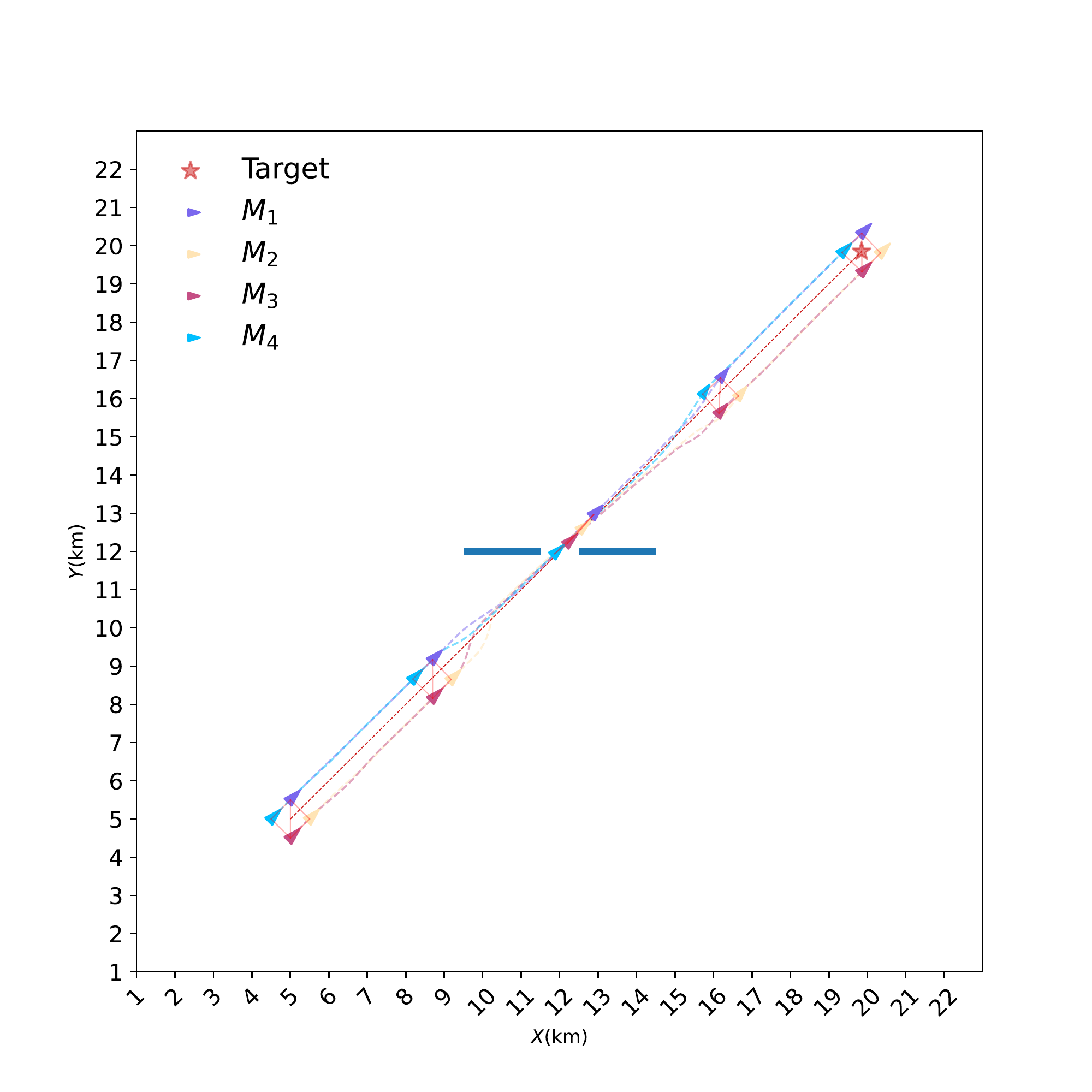}
	\caption{Trajectories of switching formation type case}
	\label{fig:switchTypeTraj}
\end{figure}

\begin{figure}[htbp]
	\centering
	\includegraphics[scale=0.4]{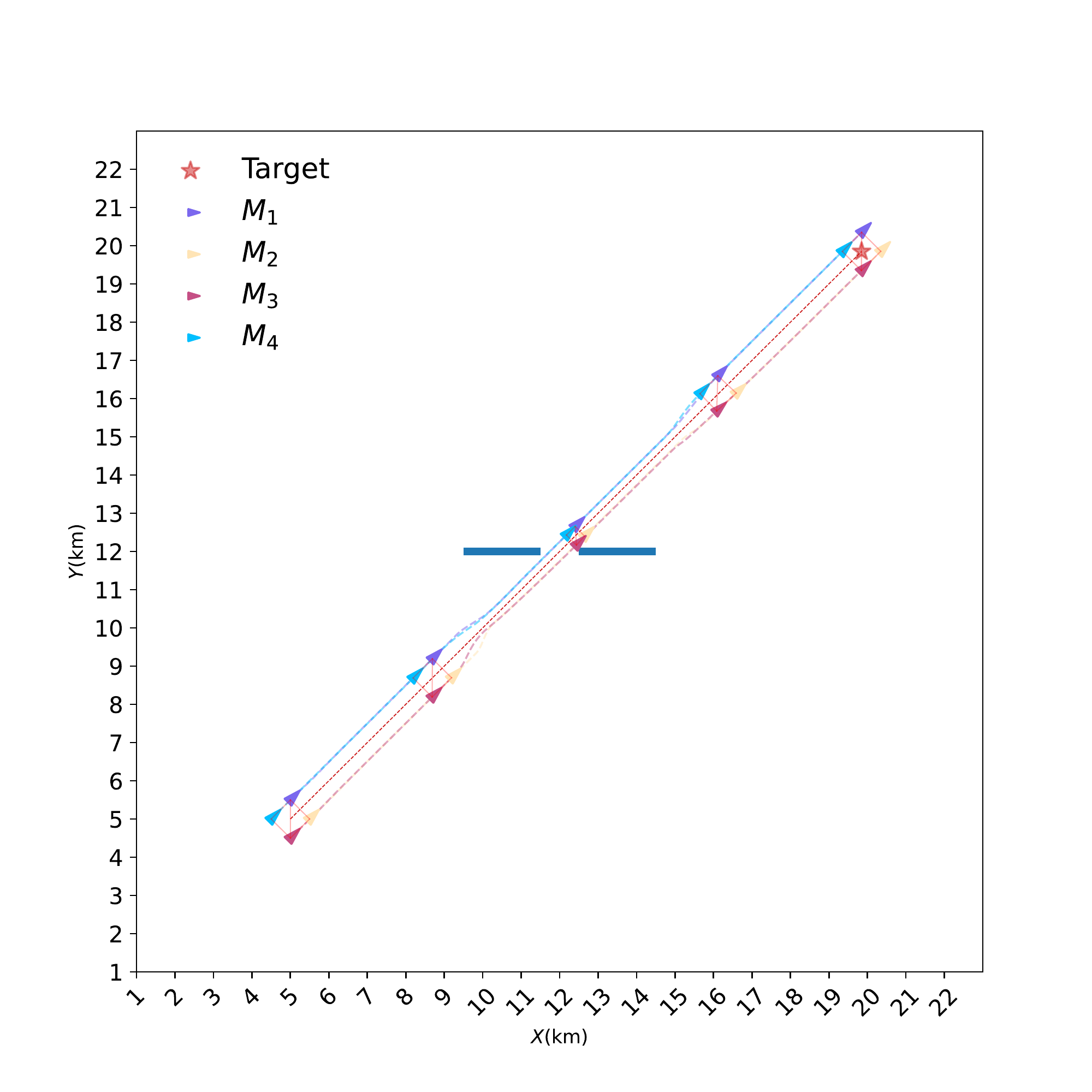}
	\caption{Trajectories of switching formation size case}
	\label{fig:switchSizeTraj}
\end{figure}

To cross the obstacle, we consider both changing the formation pattern and formation size, and the time-varying definition of the formation in both cases are as follows:
\begin{equation}
	\begin{aligned}
		\lambda^1(t) &= \begin{cases}
			\lambda^R_{(0,0.5)}, &\text{if }d_c\leqslant n\cdot l_f \\
			\lambda^R_{(-\pi/4, 0.5)}, & \text{otherwise}.
		\end{cases},\\
		\lambda^2(t) &= \begin{cases}
			\lambda^R_{(0,0.5)}, &\text{if }d_c\leqslant n\cdot l_f \\
			\lambda^R_{(0, 0.2)}, & \text{otherwise}.
		\end{cases},\\
	\end{aligned}
\end{equation}
The result trajectories are shown in \prettyref{fig:switchTypeTraj} and \prettyref{fig:switchSizeTraj}, it can be observed that when obstacles are detected, the swarm can swiftly adjust by changing the formation pattern or size, and then recover the formation quickly after going through.


\subsection{Formation control under node failure}

To verify the effectiveness of the proposed algorithm under node failure. In this experimental scenario, a swarm with six nodes are designed to pursue the reference target in a regular polygon formation, and the reference target moves in a sinusoidal fashion. At $t=20s$ during the pursuit, the cluster head node 1 and the cluster member node 4 suffer attacks and will disconnect from the other nodes in the swarm, and the remaining nodes of the swarm need to maintain their original formation and complete the formation task.

\begin{figure}[htbp]
	\centering
	\includegraphics[scale=0.4]{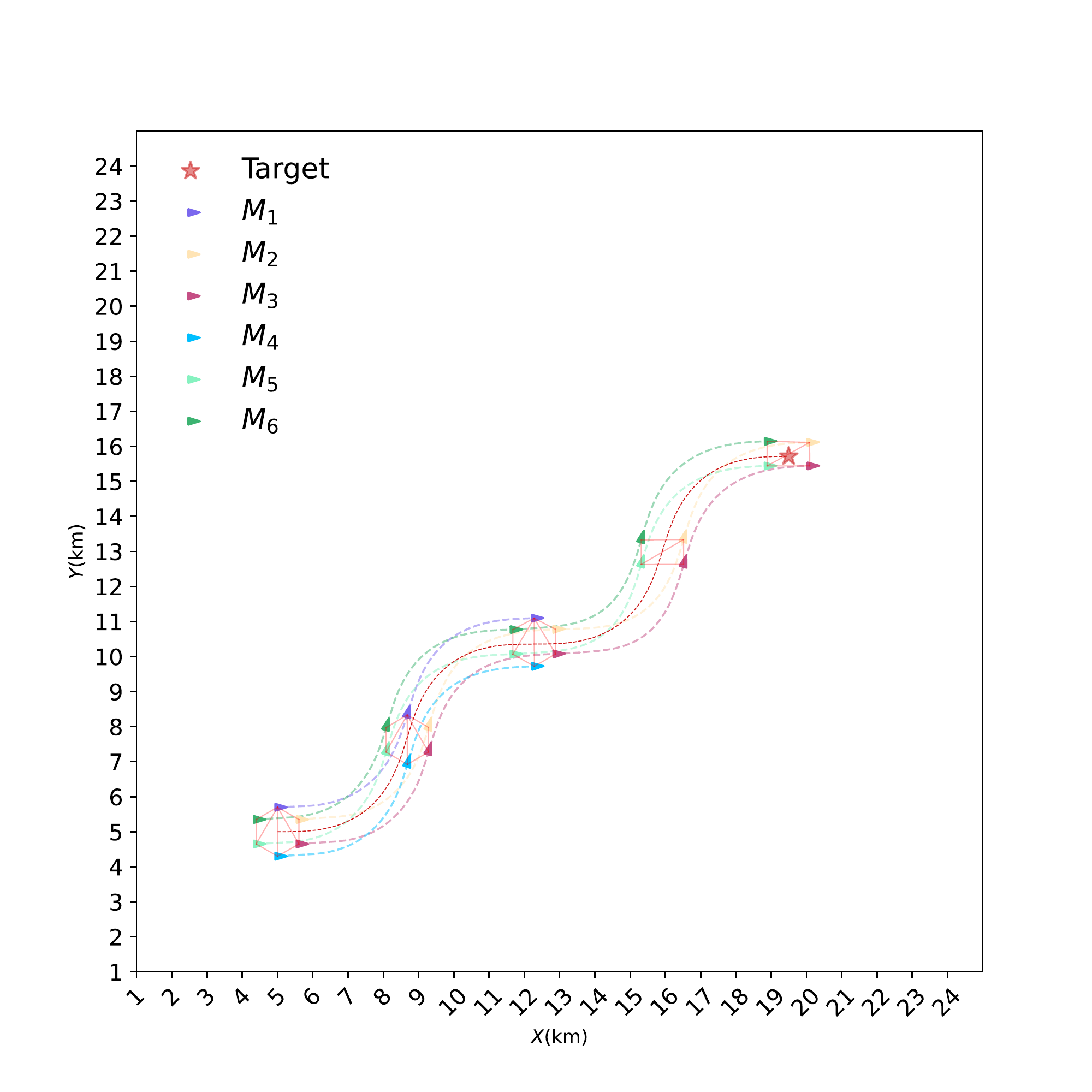}
	\caption{Trajectories of the node failure case}
	\label{fig:nodeFailTraj}
\end{figure}
\begin{figure}[htbp]
	\centering
	\includegraphics[scale=0.4]{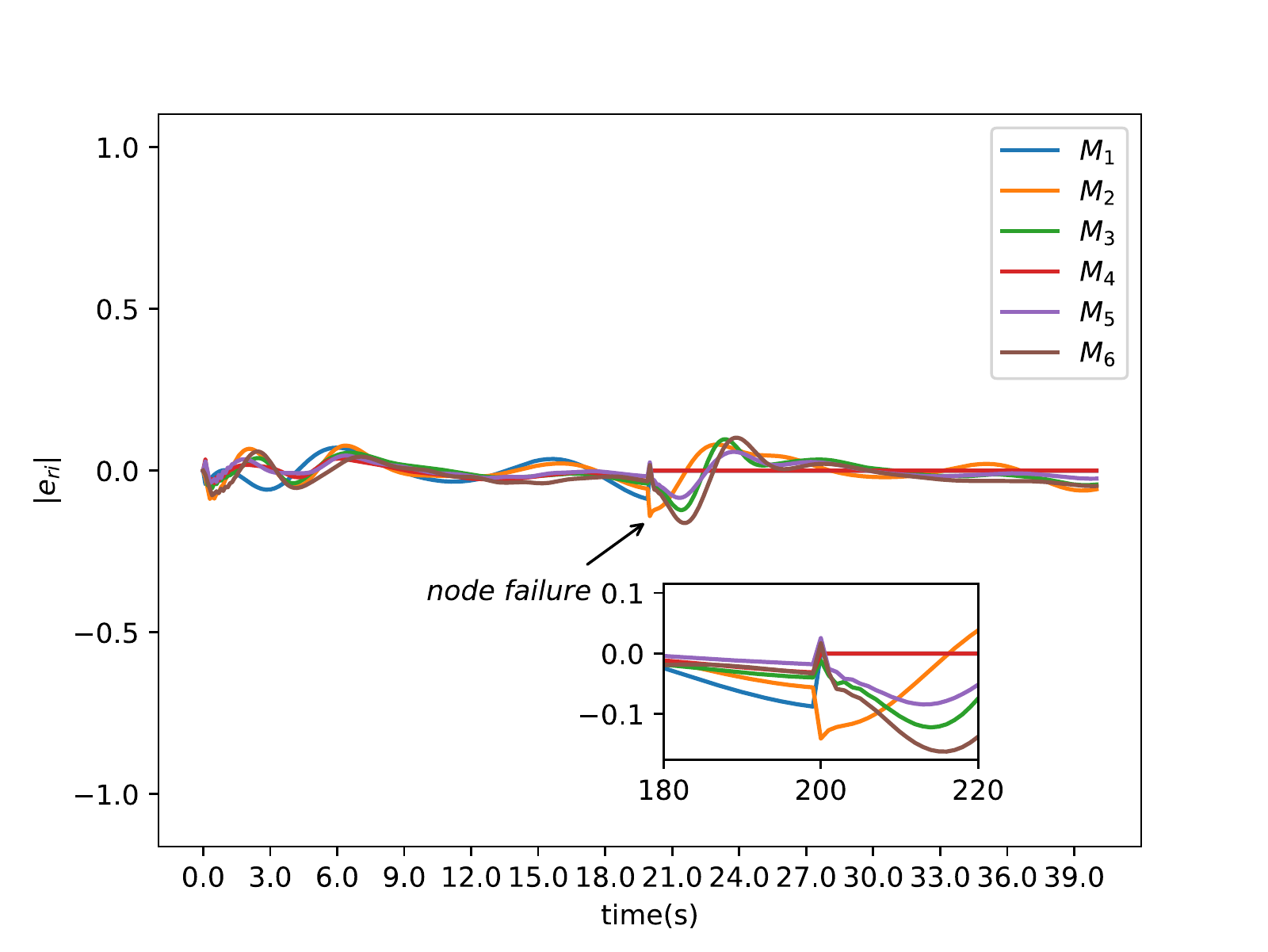}
	\caption{Resultant error curves of the node failure case}
	\label{fig:resErrNodeFail}
\end{figure}
From the results \prettyref{fig:nodeFailTraj} and \prettyref{fig:resErrNodeFail}, it can be observed that the swarm selects node 2 as the cluster head to reorganize the communication topology after the node failure and successfully maintains the original formation shape after a short fluctuation. Finally, the robustness of the proposed formation control algorithm against node failure is validated.
To investigate the effect of policy constraint on control performance, we compared the results of  the NCES-based formation control method that impose policy constraints with the one without policy constraints. In some cases such as node failure and switching formation, there is a certain chance that the algorithm will not converge, moreover, in all cases the convergence rate is generally improved by more than 20 percents with policy constraint than without it, therefore, policy constraint is essential for the training of NCES-based neural network controller.

\section{Conclusion}
\label{section:Conclusion}
In this paper, a novel distributed NCES based formation control algorithm for a second-order multi-missile system using neural network controller has been proposed. The algorithm minimizes the formation shape error and tracking error by training the optimal network controller through iterative learning, combining with a policy constraint approach to enhance the stability of the algorithm. Additionally, we have designed an adaptive topology scheme for the node failure situation, which can achieve stable communication connections at a low communication cost. We have also proposed a stable population adaptation method base on evolution path, in order to further improve the performance of the algorithm and alleviate local optimum issue. Extensive experiments demonstrated that the proposed formation control algorithm is capable of accomplishing tasks such as formation maintenance, tracking of reference trajectories and formation transformation in face of obstacles with high accuracy, and with certain robustness to cope with situations such as random initial positions and node failures. Future work may include the investigation of online evolutionary algorithms to solve the problem of unknown or stochastic system models, building on existing work.



\clearpage
\bibliographystyle{elsarticle-num} 
\bibliography{Swarm-Formation-Control-Bibliography}





\end{document}